\documentclass[prd,aps,amsmath,amssymb,reprint]{revtex4-2}

\usepackage{graphicx}
\usepackage{booktabs}
\usepackage{array}
\usepackage{xcolor}
\usepackage[breaklinks,colorlinks=true,linkcolor=blue,citecolor=blue]{hyperref}

\newcommand{\Sp}{\Sigma_p}
\newcommand{\Sr}{\Sigma_r}
\newcommand{\Ssh}{\Sigma_s}
\newcommand{\Ib}{\bar I}
\newcommand{\Qb}{\bar Q}
\newcommand{\lbt}{\bar\lambda_{\rm tid}}
\newcommand{\ec}{\epsilon_c}
\newcommand{\gm}{\tfrac{\gamma}{1-\gamma}}
\newcommand{\gp}{\tfrac{1}{1-\gamma}}
\newcommand{\cU}{\mathcal{U}}
\newcommand{\cO}{\mathcal{O}}
\newcommand{\cR}{\mathcal{R}}

\newcommand{\cA}{\mathcal{A}}
\newcommand{\cP}{\mathcal{P}}
\newcommand{\cM}{\mathcal{M}}

\newcommand{\cS}{\mathcal{S}}
\newcommand{\cN}{\mathcal{N}}
\newcommand{\cI}{\mathcal{I}}
\newcommand{\cL}{\mathcal{L}}
\newcommand{\nb}{N_0}
\newcommand{\fa}{\mathfrak{a}}
\newcommand{\fb}{\mathfrak{b}}

\begin{document}

\title{Universal Relations for Neutron Stars from Asymptotic Analysis}

\author{Syo Kamata}
\email{skamata11phys@gmail.com}
\affiliation{Department of Physics, The University of Tokyo, 7-3-1 Hongo, Bunkyo-ku, Tokyo 113-0033, Japan}

\author{Josuke Minamiguchi}
\email{minamiguchi@nt.phys.s.u-tokyo.ac.jp}
\affiliation{Department of Physics, The University of Tokyo, 7-3-1 Hongo, Bunkyo-ku, Tokyo 113-0033, Japan}

\author{Shuhei Minato}
\email{minato@nt.phys.s.u-tokyo.ac.jp}
\affiliation{Department of Physics, The University of Tokyo, 7-3-1 Hongo, Bunkyo-ku, Tokyo 113-0033, Japan}

\date{\today}

\begin{abstract}
    Dimensionless observables of neutron stars, such as the moment of inertia,
    the tidal deformability, the spin-induced quadrupole moment, and the
    compactness, satisfy the I--Love--Q and Love--$C$ universal relations to
    percent-level accuracy over a wide range of equations of state.
    We investigate analytically the origin of this insensitivity in the stellar structure equations.
    We derive asymptotic expansions of these relations directly from the
    differential equations and boundary conditions for slowly rotating, tidally
    deformed stars described by a general piecewise-polytropic equation of state.
    Although the differential equations allow several forms of the asymptotic
    expansion, we find that only one class is consistent with the observed 
    universality, and we use this class to analyze the universal relations.
    Because the observables are determined at the stellar surface, information
    about the deep-interior equation of state can enter them only through the
    parameters that survive in the asymptotic expansion at the surface.
    We find that the universal relations depend only on three parameters:
    two integration constants and one polytrope index of the outermost segment.
    By determining how these parameters deform the relations, we show that,
    throughout the region of parameter space occupied by realistic equations of
    state, the resulting deviations remain at the percent level, consistent with
    the observed accuracy of the universal relations.
    Within the same formalism, we classify violations of universality according
    to the additional degrees of freedom or input data responsible for them.
    Since the construction relies only on the underlying differential equations
    and boundary conditions, the same procedure can be applied to other systems
    once the corresponding equations and boundary conditions are specified.
\end{abstract}

\maketitle

\tableofcontents

\section{Introduction}
\label{sec:intro}
The I--Love--Q relations \cite{Yagi:2016bkt} are approximately
equation-of-state (EoS) independent relations among the dimensionless moment
of inertia, tidal deformability, and spin-induced quadrupole moment. 
For any pair of them, the relation expressing one as a function of
the other depends on the choice of realistic EoS only at the percent
level. The relations were first identified for slowly rotating, tidally
deformed stars \cite{Yagi:2013bca,Yagi:2013awa}, and have since been examined
for rapidly rotating stars
\cite{Chakrabarti:2013tca,Doneva:2013rha,Pappas:2013naa,Breu:2016ufb},
magnetized stars \cite{Haskell:2013vha}, and white dwarfs
\cite{Boshkayev:2016,Boshkayev:2018,Taylor:2019hle}. A companion relation ties
the tidal deformability to the compactness (Love--$C$) \cite{Maselli:2013mva,Yagi:2016bkt}.

The bulk properties of neutron stars,
most notably their masses and radii, depend strongly on the EoS of dense
nuclear matter, which remains uncertain
\cite{Oertel:2016bki,Lattimer:2021emm}. This strong EoS dependence also makes
neutron-star observations a valuable probe of dense nuclear matter. 
Neutron-star mass measurements from radio pulsar timing
\cite{Demorest:2010bx,Antoniadis:2013pzd,Fonseca:2016tux,Cromartie:2019kug},
mass--radius measurements from NICER pulse-profile modeling
\cite{Riley:2019yda,Miller:2019cac,Riley:2021pdl,Miller:2021qha}, and
X-ray spectroscopy of thermonuclear bursts and quiescent low-mass X-ray
binaries \cite{Ozel:2015fia,Bogdanov:2016nle}, together with
tidal-deformability constraints from gravitational-wave observations of
binary neutron-star mergers
\cite{LIGOScientific:2017vwq,LIGOScientific:2018cki,LIGOScientific:2020aai},
place increasingly stringent constraints on the EoS
\cite{Miller:2019nzo,Raaijmakers:2019dks,Essick:2019ldf,Annala:2019puf,%
Fujimoto:2021zas,Huth:2021bsp,Brandes:2022nxa,Annala:2023cwx,Legred:2024,%
Fujimoto:2024cyv}.
The insensitivity of the universal relations is therefore remarkable,
motivating an investigation of why the EoS dependence is strongly reduced
when these observables are related to one another.

The universality was originally established using a restricted set of equations
of state, but subsequent studies have tested its robustness over increasingly
broad EoS ensembles. In particular, Gaussian-process constructions provide
nonparametric ensembles that span a much wider range of dense-matter behavior
\cite{Essick:2019ldf}. The I--Love--Q relations remain the least sensitive to
the EoS among the neutron-star universal relations tested with these ensembles
\cite{Legred:2024}. Their accuracy, however, is not uniform: larger deviations
occur for relations involving the tidal deformability and for ensembles that
admit phase transitions, with systematic errors already comparable to current
statistical uncertainties \cite{Legred:2024}. These results show that the
universality is not an artifact of the original EoS sample, while also
demonstrating that its accuracy depends on the range of EoS behavior being
considered.

At the same time, the universal relations do not retain the same accuracy
under all stellar conditions. Deviations have been found for rapidly rotating
stars \cite{Doneva:2013rha}, strongly magnetized stars
\cite{Haskell:2013vha}, stars with anisotropic pressure
\cite{YagiYunes:2015aniso}, elastic phases
\cite{Lau:2017,Penner:2011,Gittins:2020}, or superfluid components
\cite{Yeung:2021}, as well as for newborn stars with unrelaxed entropy
gradients \cite{Martinon:2014} and for the second stable branch associated
with a strong first-order phase transition \cite{Benitez:2021}. The magnitude
of the violation depends on the physical setting and on how the stellar
sequence is defined. For example, much of the degradation caused by rapid
rotation is removed when sequences are compared at fixed dimensionless spin
\cite{Chakrabarti:2013tca}. These cases have largely been studied
individually, and a general criterion for determining which effects preserve
the universal relations and which lead to significant violations is still
lacking. 

Several explanations have been proposed for the origin of the universal
relations. Early arguments emphasized either the similarity of realistic EoS
at low densities or the black-hole no-hair property
\cite{Yagi:2013bca,Yagi:2013awa}. It was later shown that the approximate
self-similarity of isodensity surfaces makes the relations predominantly
sensitive to matter outside the stellar core \cite{Yagi:2014qua}. These
arguments characterize properties of the stellar solutions, but do not by
themselves identify how the suppression of EoS dependence emerges from the
underlying structure equations. More recent studies have therefore pursued
analytic derivations directly from those equations. Such
analyses, however, generally employ simplifying assumptions: a single
polytrope \cite{Yip:2017} or another model EoS
\cite{Sham:2014kea,Chan:2014tva,Chan:2015iou} over the full density range,
phenomenological density profiles such as Tolman models
\cite{Jiang:2019vmf,Jiang:2020uvb,Lowrey:2024anh,Katagiri:2025}, or an
expansion about the Newtonian limit \cite{Yip:2017,Saes:2024}. 
They therefore do not fully capture the freedom of a realistic EoS, in which
regions of different stiffness are connected across interfaces. The stiffness
of each segment and the density at which one segment gives way to the next
can vary independently, generating a correspondingly broad range of stellar
interior structures. The crucial question is therefore how much information about this
EoS-dependent interior structure reaches the stellar surface, where the
observables are determined.

We address this problem using asymptotic analysis, which provides a systematic
treatment of the nonlinear differential equations when closed-form
solutions are unavailable. We consider stars described by piecewise
polytropic EoS \cite{Read:2008iy} within the standard perturbative treatment of
slow rotation and static tides
\cite{Hartle:1967he,Hartle:1968si,Hinderer:2007mb,Damour:2009vw}.
The admissible asymptotic expansions are determined by the differential
equations and their boundary conditions, without assuming a particular form
of the solutions. This allows us to identify the independent data carried
between successive stellar layers. The resulting asymptotic structure and
its generators constrain the qualitative form of the universal relations
even before all expansion coefficients are determined.

The present analysis differs from earlier analytic treatments in two respects.
First, it retains the full coupled system in general relativity without
restricting the solution to a special class. Second, the asymptotic
construction is performed independently of any particular choice of
observables. It therefore applies to any relation between observables governed
by the same system of differential equations, including the I--Love--Q and
Love--$C$ relations.

Our main results are summarized as follows. We construct the asymptotic
expansions of the observables explicitly and find that they are controlled by
three parameters: the two integration constants $(\Sp,\Sr)$ transmitted
outward by each segment and the polytrope index $\gamma$ of the outermost
segment. The sensitivity of a relation between two observables is therefore
determined by how variations of these parameters displace the corresponding
stellar sequence in the observable plane.
A variation of $\Sp$ moves the sequence along the universal curve and
hence does not generate a deviation from the relation. The remaining
sensitivity is controlled by $\gamma$ and $\Sr$. The dependence on $\gamma$
is weak over the range relevant to realistic EoS. The effect of $\Sr$ depends
on its magnitude. For moderate $\Sr$, its contribution to the observables is
suppressed with increasing compactness as $C^{-3/2}$. Since $(\Sp,\Sr)$ are
the only quantities in the expansion that retain information about the inner
EoS, the suppression of their contributions implies a progressive loss of
sensitivity to the deep interior. We refer to this mechanism as
\emph{memory loss}.
For sufficiently large $\Sr$, however, this suppression is no longer
effective. Such large values can arise, for example, from a strong first-order phase
transition with a large latent heat. In this regime the $C^{-3/2}$
suppression of the $\Sr$-dependent contribution is
insufficient to keep the
stellar sequence close to the universal curve, and the universality is lost.

This paper is organized as follows. Section~\ref{sec:setup} introduces the
stellar structure equations, the family of EoS, and the observables.
Section~\ref{sec:general} develops the asymptotic expansion, and
Sec.~\ref{sec:which} identifies the parameters governing the universal
relations. Sections.~\ref{sec:gamma} and \ref{sec:Sr} examine the dependence
on the polytrope index $\gamma$ and the integration constant $\Sr$,
respectively. Section~\ref{sec:disc} discusses the scope and limitations of
the framework, and Sec.~\ref{sec:concl} summarizes our conclusions.
Throughout this paper, we use geometrized units $G=1=c$.

\section{Formulation}
\label{sec:setup}

This section introduces the coupled differential equations and their boundary
conditions, the family of equations of state, and the definitions of the
observables used in the universal relations.

\subsection{Coupled differential equations}
\label{sec:eqs}

We treat the matter as a perfect fluid and write the metric as
$ds^2=-e^{\nu}dt^2+e^{\lambda}dr^2+r^2d\Omega^2$ with
$e^{\lambda}=(1-2m/r)^{-1}$. The background variables $(p,m,\nu)$ obey the
Tolman--Oppenheimer--Volkoff (TOV) equations~\cite{Tolman:1939jz,Oppenheimer:1939ne}
\begin{align}
    \frac{dp}{dr}&=-\frac{(p+\epsilon)(m+4\pi r^3 p)}{r(r-2m)},
    \nonumber\\
    \frac{dm}{dr}&= 4\pi r^2\epsilon,
    \qquad
    \frac{d\nu}{dr}=2\,\frac{4\pi r^3 p+m}{r(r-2m)} .
    \label{eq:TOV}
\end{align}
Slow rotation and a static tidal deformation are described by perturbations
about this background. We denote the $O(\Omega)$ perturbation
by $\omega_1$, the $O(\Omega^2)$ rotational $\ell=2$ perturbations by
$(K_2,h_2)$, and the static tidal $\ell=2$ perturbation by $H_0$. They satisfy
\cite{Hartle:1967he,Hartle:1968si,Hinderer:2007mb,Damour:2009vw,%
Yagi:2013awa}
\begin{align}
    &\frac{d^2\omega_1}{dr^2}+4\,\frac{1-\pi r^2(p+\epsilon)e^{\lambda}}{r}\frac{d\omega_1}{dr}
    \nonumber\\
    &\qquad-16\pi(p+\epsilon)e^\lambda \omega_1=0,\label{eq:omegadiff}\\
    &\frac{dK_2}{dr}=-\frac{dh_2}{dr}+\frac{r-3m-4\pi r^3 p}{r^2}e^{\lambda}h_2
    \nonumber\\
    &\qquad+\frac{r-m+4\pi r^3 p}{r^3}e^{2\lambda}m_2,\label{eq:k2diff}\\
    &\frac{dh_2}{dr}=-\frac{r-m+4\pi r^3 p}{r}e^\lambda\frac{dK_2}{dr}
    \nonumber\\
    &\qquad+\frac{3-4\pi r^2(p+\epsilon)}{r}e^{\lambda}h_2+\frac{2}{r}e^\lambda K_2
    \nonumber\\
    &\qquad+\frac{1+8\pi r^2 p}{r^2}e^{2\lambda}m_2
      +\frac{r^3}{12}e^{-\nu}\Big(\frac{d\omega_1}{dr}\Big)^2
    \nonumber\\
    &\qquad-\frac{4\pi(p+\epsilon)r^3\omega_1^2}{3}e^{-\nu+\lambda},\label{eq:h2diff}\\
    &\frac{d^2 H_0}{dr^2}+\frac{d H_0}{dr}\Big\{\frac{2}{r}
      +e^{\lambda}\Big[\frac{2m}{r^2}+4\pi r(p-\epsilon)\Big]\Big\}
    \nonumber\\
    &\quad+H_0\Big[-\frac{6 e^{\lambda}}{r^2}
      +4\pi e^{\lambda}\Big(5\epsilon+9p+\frac{p+\epsilon}{c_s^2}\Big)
    \nonumber\\
    &\qquad\qquad-\Big({\frac{d\nu}{dr}}\Big)^2\Big]=0.
      \label{eq:H0diff}\\
    &m_2=-re^{-\lambda}h_2+\frac{1}{6}r^4e^{-\nu+\lambda}\Big[re^{-\lambda}\Big(\frac{d\omega_1}{dr}\Big)^2\nonumber\\
    &\qquad\qquad+16\pi r\omega_1^2 (p+\epsilon)\Big]\label{eq:m2}
\end{align}
where $c_s^2 =dp/d \epsilon$ and $e^{-\lambda}=1-2(m/{r})$. 

\subsection{Equation of state and flow variable}
\label{sec:eos}

We represent the equation of state by a piecewise-polytropic form 
\cite{Read:2008iy, Damour:2009vw} as
\begin{equation}   \epsilon(p)=K_i\,p^{\gamma_i},\qquad p_i\le p<p_{i+1},
    \label{eq:eos}
\end{equation}
with $i=0,1,\dots,N$. For interfaces without a first-order phase transition, the matching constants are chosen such that
\begin{equation}
    K_i\,p_i^{\gamma_i}=K_{i-1}\,p_i^{\gamma_{i-1}} .
    \label{eq:eoscont}
\end{equation}
For each segment, we impose
\begin{equation}
    \gamma_i>0,\qquad
    c_s^{2}=\frac{dp}{d\epsilon}=\frac{p}{\gamma_i\,\epsilon}\le1 .
    \label{eq:causal}
\end{equation}
The speed of sound $c_s$ appearing in Eq.~\eqref{eq:H0diff} is determined by
Eq.~\eqref{eq:eos}. That is, the $\ell=2$ fluid perturbation uses the same equation of state as the background, which amounts to imposing
\begin{equation}
\delta\epsilon=\frac{d\epsilon}{dp}\,\delta p .
    \label{eq:pertEOS}
\end{equation}

In solving Eqs.~\eqref{eq:TOV}--\eqref{eq:H0diff}, we use a dimensionless variable, $c=m/r$, as a flow parameter. This satisfies
\begin{align}
    \frac{dc}{dr}=\frac{4\pi r^2\epsilon-c}{r} .
    \label{eq:flowvar}
\end{align}
Note that the variable $c$ vanishes at the center of the neutron star, and equals the compactness $C=M/R$ at
the surface. 
Because the observables are read off only at the surface, $C$ uniquely specifies the point at which the asymptotic expansions must be evaluated. In addition, the flow parameter $c$ is bounded by the black-hole limit, $0\le c<1/2$, which makes it a natural parameter for organizing the
asymptotic expansion. 
As a measure of relativistic effects, we use
the ratio of the pressure to energy density at the stellar center
\begin{equation}
    q_c\equiv\frac{p_c}{\ec},
    \label{eq:qc}
\end{equation}
whose limit $q_c\to0$ is the Newtonian limit.
Rewriting Eqs.~\eqref{eq:TOV}--\eqref{eq:m2} in these variables gives
\begin{align}
    &\frac{dp}{dc} = -\frac{({\epsilon} + p)(c + 4 \pi r^2 p)}{(4 \pi {\epsilon} r^2 - c)(1-2 c)},
      \qquad
      \frac{dr}{dc} = \frac{r}{4 \pi {\epsilon} r^2 - c}, \label{eq:dpdc}\\
    &\frac{d \nb}{d c} = -2\, \frac{4 \pi r^2 p+c}{(4 \pi {\epsilon} r^2 - c)(1-2c)}\, \nb,
      \qquad \nb \equiv e^{-\nu}, \label{eq:dNdc}\\
    &\frac{dH_0}{dc} = \frac{\beta r}{4 \pi {\epsilon} r^2 - c},\label{eq:dH0dc}\\
    &\frac{d\beta}{dc}=\frac{2 H_0 r}{(4 \pi {\epsilon} r^2 - c)( 1 - 2 c)}
      \biggl[ - 2 \pi \Big( 5 {\epsilon} + 9 p + \frac{{\epsilon} + p}{c_s^2} \Big)
    \nonumber\\
    &\qquad\qquad + \frac{3}{r^2} + \frac{2 ( c + 4 \pi r^2 p)^2}{r^2 ( 1 - 2 c)} \biggr]
    \nonumber\\
    &\quad+ \frac{2 \beta\left[ -1 + c + 2 \pi r^2 ({\epsilon} - p) \right]}{(4 \pi {\epsilon} r^2 - c)( 1 - 2 c)},\label{eq:dbetadc}\\
    &\frac{d \omega_1}{dc} = \frac{\phi r}{4 \pi {\epsilon} r^2 - c},
    \label{eq:domdc} \\
    &\frac{d\phi}{dc}=-\frac{4\,\phi}{4 \pi {\epsilon} r^2 - c}
      +\frac{4 \pi ({\epsilon} + p)\, r\,(r\phi+4\,\omega_1)}{(4 \pi {\epsilon}r^2 - c) (1-2c)}, \label{eq:dphidc} \\
    &\frac{d {K}_2}{d c} = - \frac{d {h}_2}{d c} + \frac{1}{4 \pi {\epsilon} r^2 -c}
      \Big[\frac{1 - 3 c - 4 \pi r^2 p}{1-2c} {h}_2
    \nonumber\\
    &\qquad+ \frac{1 - c + 4 \pi r^2 p}{r(1-2c)^2} \mu_2\Big], \label{eq:dk2dc} \\
    &\frac{d { h}_2}{d c} = - \frac{1 - c + 4 \pi r^2 p}{1-2c} \frac{d {K}_2}{d c}
      + \frac{1}{4 \pi {\epsilon} r^2 - c}\Big[\frac{r^4}{12} \nb \phi^2
    \nonumber\\
    &\qquad+\frac{(3{h}_2+2{K}_2) - 4 \pi r^2 ({\epsilon}+p)
      ({h}_2+\frac{r^2}{3}\nb \omega_1^2)}{1-2c}
    \nonumber\\
    &\qquad
      +\frac{1 + 8 \pi r^2 p}{r(1-2c)^2} \mu_2 \Big], \label{eq:dh2dc}\\
    &\mu_2 = (1 - 2 c)\Big[ -r{ h}_2
    \nonumber\\
    &\qquad+ \frac{r^5}{6}\nb
      \left\{ (1-2 c) \phi^2 + 16 \pi \omega_1^2 ({\epsilon}+p )\right\}\Big] .
      \label{eq:mu2}
\end{align}
Here we introduce $\beta=dH_0/dr$, $\phi=d\omega_1/dr$, and $\mu_2=m_2$.
We collect the variables used in the present analysis as
\begin{align}
    \cO\in\{\,&p_c-p,\ r,\ H_0,\ \beta,\ \omega_1,\ \phi,\ \nb,
    \nonumber\\
    &K_2^{S},\ h_2^{S},\ K_2^{H},\ h_2^{H}\,\},
    \label{eq:Oset}
\end{align}
where the superscripts $S$ and $H$ denote the particular and homogeneous solutions of Eqs.~\eqref{eq:dk2dc} and \eqref{eq:dh2dc}, respectively.

Throughout this paper, we consider the setup described above, but generalizations of the equation of state in Eq.~\eqref{eq:eos} and violation of Eq.~\eqref{eq:pertEOS} are discussed in Secs.~\ref{sec:eosgen} and~\ref{sec:shape}, respectively.

\subsection{Boundary conditions and observables}
\label{sec:obs}

The boundary-value problem for Eqs.~\eqref{eq:dpdc}--\eqref{eq:dh2dc} is completed by specifying regularity at the
center and matching to the exterior solution at the stellar surface.
The regularity of the background variables at the center requires
\begin{align}
    &m(r)=\tfrac{4\pi}{3}\ec r^3+O(r^5),
    \nonumber\\
    &p(r)=p_c-\tfrac{2\pi}{3}(\ec+p_c)(\ec+3p_c)\,r^2+O(r^4),
    \label{eq:centralbg}
\end{align}
and the perturbation variables, $\omega_1, H_0, h_2$, and $K_2$, behave as
\begin{align}&\omega_1(r)=\omega_c\left[1+O(r^2)\right],\qquad
    H_0(r)=a_0\,r^2+O(r^4),\nonumber\\
    &h_2(r)=A\,r^2+O(r^4),\qquad K_2(r)=-A\,r^2+O(r^4) .
    \label{eq:centralpert}
\end{align}
The regular central solution in each perturbation sector contains one free
constant. The constants $\omega_c$ and $a_0$ only set the normalization of
the corresponding linear solutions and therefore cancel from the observables.
In contrast, the constant $A$ determines the relative contribution of the regular
homogeneous $\ell=2$ solution to the rotationally sourced particular solution
and is fixed by the surface matching, as discussed in
Sec.~\ref{sec:layer}.

The surface is defined by $p(R)=0$, with $M=m(R)$. The definition of compactness is $C\equiv M/R$.
The exterior solution of Eqs.~\eqref{eq:domdc} and~\eqref{eq:dphidc} is specified as $\omega_1^{\rm ex}=\Omega-2J/r^3$ by two constants $\Omega$ and $J$. The 
continuity of $\omega_1$ and $d\omega_1/dr$ at the stellar surface gives
\begin{equation}
    J=\frac{R^4}{6}\left.\frac{d\omega_1}{dr}\right|_{r=R},\qquad I=\frac{J}{\Omega} .
    \label{eq:Jdef}
\end{equation}
Matching the rotational quadrupole to the exterior $\ell=2$ solution in the same
way yields the quadrupole moment \cite{Hartle:1968si} as
\begin{equation}
    Q=-\frac{J^2}{M}-\frac{8}{5}\,K_Q\,M^3,
    \label{eq:Qdef}
\end{equation}
where the constant $K_Q$ is fixed by the matching between internal and external solutions~\cite{Yagi:2013bca}. For the computation of $Q$, a reduced system corresponding to
Eqs.~\eqref{eq:k2diff} and~\eqref{eq:h2diff} was recently formulated in
Ref.~\cite{Kyutoku:2025zud}.

For the tidal deformation, the exterior solution can be written with associated
Legendre functions as
$H_0^{\rm ex}(r)=c_1Q_2^{2}(r/M-1)+c_2P_2^{2}(r/M-1)$, and the tidal Love number
$k_2^{\rm tid}$ is determined by the ratio $c_1/c_2$. This ratio follows from the
surface matching conditions as
\begin{align}
    &H_0^{\rm in}(R)=H_0^{\rm ex}(R),
    \nonumber\\
    &\frac{dH_0^{\rm in}}{dr}(R)-\frac{4\pi R^2\epsilon(R^-)}{M}\,H_0^{\rm in}(R)
    =\frac{dH_0^{\rm ex}}{dr}(R),
    \label{eq:H0match}
\end{align}
where the second term on the left-hand side represents the jump of $dH_0/dr$
that appears when the energy density drops discontinuously to zero at the
surface. The matching conditions show that the ratio $c_1/c_2$ depends on the
interior solution only through the single combination
\begin{equation}
    y\equiv\frac{R\,H_0'(R^-)}{H_0(R)}-\frac{4\pi R^3\epsilon(R^-)}{M},
    \label{eq:ydef}
\end{equation}
and $k_2^{\rm tid}$ follows in closed form from $y$ and $C$
\cite{Hinderer:2007mb,Damour:2009vw,Postnikov:2010yn}:
\begin{equation}
    k_2^{\rm tid}=\frac{8}{5}\,
    \frac{C^{5}\,(1-2C)^{2}\,\bigl[2+2C(y-1)-y\bigr]}{D_k},
    \label{eq:k2tid}
\end{equation}
with
\begin{align}
    D_k&=2C\bigl[6-3y+3C(5y-8)\bigr]
    \nonumber\\
    &\quad+4C^{3}\bigl[13-11y+C(3y-2)+2C^{2}(1+y)\bigr]
    \nonumber\\
    &\quad+3(1-2C)^{2}\bigl[2-y+2C(y-1)\bigr]\log(1-2C) .
    \label{eq:k2tidD}
\end{align}
As $C\to0$, $k_2^{\rm tid}\to\frac12(2-y)/(3+y)$, recovering the Newtonian Love
number.

Following Refs.~\cite{Yagi:2013bca,Yagi:2013awa}, we define the observables as
\begin{equation}
    \Ib\equiv\frac{I}{M^3},\qquad
    \Qb\equiv-\frac{Q\,M}{I^2\Omega^2},\qquad
    \lbt\equiv\frac{2}{3}\frac{k_2^{\rm tid}}{C^5} .
    \label{eq:observables}
\end{equation}
Equations~\eqref{eq:Jdef}--\eqref{eq:k2tid} show that all three observables
are determined entirely by quantities at the stellar surface and in the
exterior solutions.

For a fixed equation of state, varying the central pressure $p_c$ generates a
one-parameter sequence of equilibrium stars and hence a curve in the space of
observables,
\begin{equation}
    p_c\;\longmapsto\;\bigl(\Ib(p_c),\;\Qb(p_c),\;\lbt(p_c)\bigr) .
    \label{eq:curve}
\end{equation}
In the following section, we determine how the equation of state enters this
curve through the general structure of the solutions.

\section{Asymptotic analysis and the three segment parameters}
\label{sec:general}

The curve \eqref{eq:curve} depends on the equation of state through the
stellar structure equations. For a piecewise-polytropic equation of state, an
inner segment affects the outer solution only through the quantities passed
across the interfaces of polytrope segments. To identify these quantities, 
we construct the general
solution of Eqs.~\eqref{eq:dpdc}--\eqref{eq:mu2} as an asymptotic series and
impose the boundary conditions of Sec.~\ref{sec:obs}.

\subsection{Segment-wise general solution and the three parameters}
\label{sec:layer}

\begin{figure*}[t]
\centering
\includegraphics[width=0.72\textwidth]{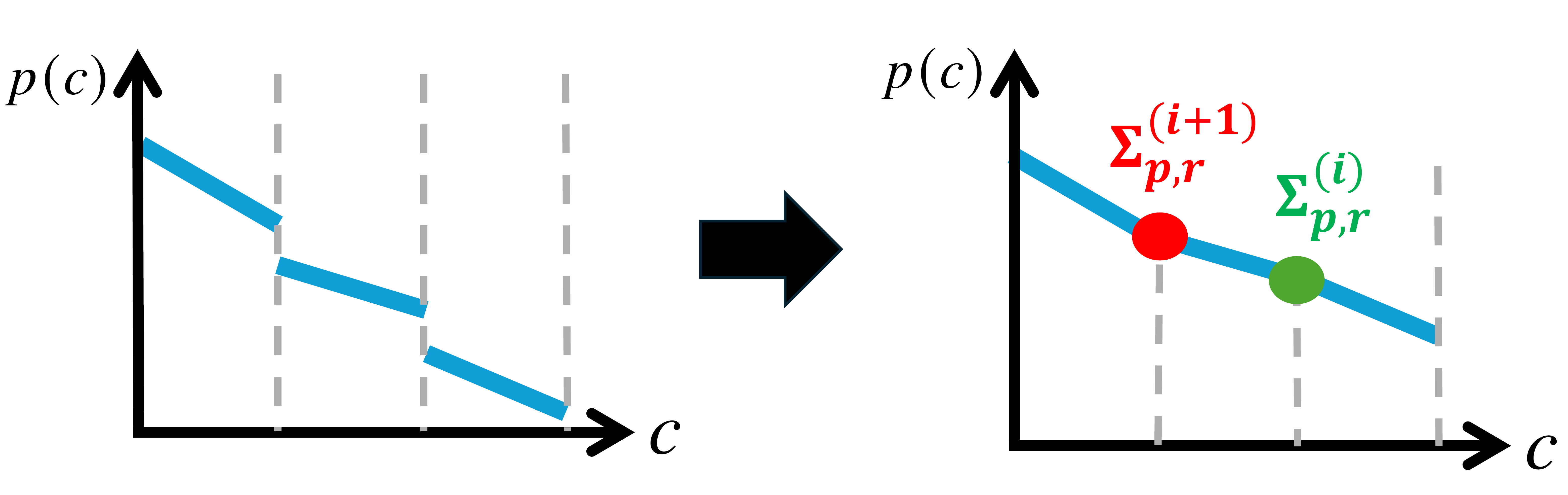}
\caption{The two constants of a segment. 
Left: if each segment were continued inward as the regular solution generated by its own equation of state, the resulting $p(c)$ curves would differ and would fail to meet at the interfaces (dashed). 
Right: the physical solution is continuous; the offset of
segment $i$ from its regular solution is recorded by the pair
$(\Sp^{(i)},\Sr^{(i)})$ of Eq.~\eqref{eq:Sdef}, fixed at the interface where
the segment begins. The constants accumulate outward and are the only data an
inner segment transmits outward. The same matching also makes $r(c)$ continuous.}
\label{fig:layers}
\end{figure*}

\subsubsection{Counting the free constants}
Equations~\eqref{eq:dNdc}--\eqref{eq:mu2} are linear,
and each linear perturbation sector associated with $N_0$, $\omega_1, H_0, h_2$, and $K_2$ carries one overall constant, $\sigma_{N_0}$, $\sigma_{\omega_1}$, $\sigma_{H_0}$, $\sigma_{h_2}$, and $\sigma_{K_2}$. Therefore, demanding that these overall constants drop out of the equations gives
\begin{align}
    &\sigma_{H_0}=\sigma_{\beta},\qquad \sigma_{\omega_1}=\sigma_{\phi},\qquad
    \sigma_{K_2^{H}}=\sigma_{h_2^{H}},
    \nonumber\\
    &\sigma_{K_2^{S}}=\sigma_{h_2^{S}}=\sigma_{\nb}\,\sigma_{\omega_1}^{2}. \label{eq:signorm}
\end{align}
The constants $\sigma_{\omega_1}$ and $\sigma_{H_0}$ only fix the
normalization and therefore drop out of the observables. The remaining constants are
fixed by central regularity and surface matching
\cite{Hartle:1967he,Hartle:1968si,Hinderer:2007mb,Hinderer:2009ca,%
Damour:2009vw}.

The background variables $(p,m)$ are different: unlike the perturbation sectors, the first two equations in
Eq.~\eqref{eq:TOV} carry two integration constants per segment. In the innermost segment these are fixed by regularity and the choice of central pressure, but they are generally nonzero in
subsequent segments. 
It is these integration constants that carry the interior information outward.
The coefficients of
Eqs.~\eqref{eq:dNdc}--\eqref{eq:mu2} involve the segment only through
$\epsilon(p)$, and remain bounded across an interface at which $p$ and $r$
are continuous.
The perturbation variables are continuous through it
and the matching at the interfaces introduces no additional constants of its own, as in
the treatment of density discontinuities of
Refs.~\cite{Damour:2009vw,Postnikov:2010yn}. 
The general solution in the $i$-th polytrope segment therefore depends only on the polytrope index of that segment
$\gamma_i$, the central pressure and the two integration constants of $(p,m)$.

\subsubsection{The two constants and their meaning}

To determine the integration constants of the $(p,m)$ sector, we expand the
solution in the $i$-th segment about the center of the corresponding regular
solution, where $m=0$ and $(\epsilon,p)=(\epsilon_c,p_c)$. The TOV equations
then give
\begin{align}
    m(r)&=\frac{4\pi}{3}\ec r^3+\delta m ,
    \nonumber\\
    p(r)&=p_c-\delta p-\frac{2\pi}{3}(\ec+p_c)(\ec+3p_c)\,r^2
    \nonumber\\
    &\quad+(\ec+p_c)\frac{\delta m}{r} , \label{eq:mexp}
\end{align}
where $\delta m$ and $\delta p$ denote the mass offset and central-pressure
shift, respectively. We define the corresponding dimensionless constants by
\begin{equation}
    \Sr\equiv\sqrt{3\pi\ec}\;\delta m,\qquad
    \Sp\equiv\frac{\delta p}{\ec} . \label{eq:Sdef}
\end{equation}
The pair $(\Sp^{(i)},\Sr^{(i)})$ specifies the solution in the $i$-th
segment relative to its associated regular solution, as illustrated in
Fig.~\ref{fig:layers}. It vanishes
in the innermost segment and accumulates outward across successive
interfaces. The pair for the outermost segment, $i=0$, enters the solution
at the stellar surface.  We suppress the segment label below unless it is needed.

At leading order, the shift $\delta p$ can be absorbed into a redefinition
of $p_c$ within the regular family, thereby removing $\Sp$~\footnote{This
redefinition also rescales $\Sr$ by $1+O(\Sp)$, which is a second-order
effect.}.
We therefore neglect $\Sp$ in what follows. The validity of this
approximation is discussed in Sec.~\ref{sec:sigmap}.

In contrast, $\Sr$ multiplies the singular branch and enters
Eq.~\eqref{eq:mexp} already at leading order. Immediately outside an
interface at $r=r_t$, the enclosed mass differs by $\delta m$ from that of
the regular solution of the outer segment. 
With
$\epsilon_+\equiv\epsilon(r_t^+)$, Eq.~\eqref{eq:Sdef} gives
\begin{align}
    \Sr&=\sqrt{3\pi\,\epsilon_+}\;
    \left[\,m(r_t)-\frac{4\pi}{3}\,\epsilon_+\,r_t^{3}\,\right]
    \nonumber\\
    &=\sqrt{3\pi\,\epsilon_+}\;\frac{4\pi}{3}\,r_t^{3}
    \left[\,\bar\epsilon(r_t)-\epsilon_+\,\right] ,
    \label{eq:Srread}
\end{align}
where $\bar\epsilon(r_t)\equiv3m(r_t)/(4\pi r_t^3)$ is the mean density
inside the interface. Thus, $\Sr$ measures the excess or deficit of the
enclosed mass relative to the regular solution of the outer segment, with
its sign determined by $\bar\epsilon(r_t)-\epsilon_+$.
This relation should be understood as a leading-order interpretation of $\Sr$, not a definition. 
Its definition is given separately in Eq.~\eqref{eq:Sdef}. This relation clarifies
the physical origin and parametric scaling of $\Sr$, but is not intended to
provide quantitative estimates for relativistic stars.

Finally, we express the solution in terms of the flow variable $c$. Substituting
Eq.~\eqref{eq:mexp} into Eq.~\eqref{eq:flowvar} and solving for $r$ gives
\begin{equation}
    r(c)=\left[\frac{3c}{4\pi\ec}\right]^{1/2}
        \left[1-\frac{1}{3}\frac{\Sr}{c^{3/2}}+O(\Sr^2)\right] ,
    \label{eq:rofc}
\end{equation}
and Eq.~\eqref{eq:qc} then gives
\begin{align}
    p(c)=p_c\biggl[1-c\biggl\{&\frac{(1+q_c)(1+3q_c)}{2q_c}
        +\frac{1}{q_c}\frac{\Sp}{c}
    \nonumber\\
        &-\frac{(1+q_c)^2}{q_c}\frac{\Sr}{c^{3/2}}
        +O(\Sr^2)\biggr\}\biggr] .
    \label{eq:pofc}
\end{align}
The two constants enter the solution through the combinations
\begin{equation}
    \sigma_p(c)\equiv\frac{\Sp}{c},\qquad
    \sigma_r(c)\equiv\frac{\Sr}{c^{3/2}} .
    \label{eq:sigmadef}
\end{equation}
These combinations provide the natural variables for the expansion, since
each power of $\Sp$ and $\Sr$ appears together with the corresponding powers
$c^{-1}$ and $c^{-3/2}$. Therefore, we use $\sigma_p$ and $\sigma_r$ as expansion
parameters below. 

\subsection{The expansion basis and the segment solution}
\label{sec:series}
\label{sec:cexp}

\subsubsection{Roadmap for constructing the asymptotic expansion} \label{sec:roadmap}

To construct the expansion, we first determine which monomials in $c$, $\Sp$,
and $\Sr$ can appear. The allowed monomials must form a set that is closed
under all operations in Eqs.~\eqref{eq:dpdc}--\eqref{eq:mu2}. Once such a
set is specified, substituting the series into the equations
and collecting terms with the same monomial then gives an algebraic equation
for each monomial, and the coefficients are determined recursively.

Indeed, this set can be determined from $p$ and $r$.
The right-hand sides of Eqs.~\eqref{eq:dNdc}--\eqref{eq:dphidc} involve only
$p$, $r$, $\epsilon(p)$, and variables determined within the same system. Therefore, 
the expansion bases of $\nb$, $H_0$, $\beta$, $\omega_1$, and $\phi$ are
generated by the same set of monomials as those of $p$ and $r$.
Once these expansions are determined, the source terms in
Eqs.~\eqref{eq:dk2dc}--\eqref{eq:mu2} are fixed as well, and the same
monomial set determines the expansions of $K_2^{S}$, $h_2^{S}$,
$K_2^{H}$, and $h_2^{H}$.

The construction has two steps. First, we determine the leading-order
background solutions for $p$ and $r$ while retaining the two integration
constants explicitly. This step has already been carried out in
Eqs.~\eqref{eq:rofc} and \eqref{eq:pofc}.
Second, this solution is substituted into Eq.~\eqref{eq:dpdc}, and the
operations on its right-hand side are applied recursively. Any new monomial
generated in this process must be included in the expansion. Repeating the
procedure until no additional monomials are generated determines the closed
set required at all orders. 
Since all higher-order monomials are generated recursively from the
leading-order solution of $(p,m)$, the choice of expansion basis is already
encoded in its structure. The different bases arise from the possible
orderings of the three expansion parameters $c$, $\Sp$, and $\Sr$.
We therefore begin by classifying these orderings at leading order.

\subsubsection{The triple expansion and its six orderings}
\label{sec:orderings}

The segment solution involves three expansion parameters, $c$, $\Sp$, and
$\Sr$. For a multivariable expansion, the resulting basis depends on the
order in which these parameters are expanded. The six possible orderings are
\begin{equation}
\begin{aligned}
&\text{(A)}:\ \Sr\to\Sp\to c,\qquad
&&\text{(B)}:\ \Sp\to\Sr\to c,\\
&\text{(C)}:\ \Sr\to c\to\Sp,\qquad
&&\text{(D)}:\ \Sp\to c\to\Sr,\\
&\text{(E)}:\ c\to\Sp\to\Sr,\qquad
&&\text{(F)}:\ c\to\Sr\to\Sp,
\end{aligned}
\label{eq:orderings}
\end{equation}
where the leftmost parameter is expanded first.
Our purpose is to determine which of these orderings lead to genuinely
different asymptotic expansions of the stellar solution. We first determine
the leading-order segment solution in each ordering. This solution still
contains $q_c$ as an undetermined parameter, which must be fixed to obtain a
stellar solution. Second, we impose the surface condition on our solution. The surface condition $p(C)=0$ provides the required
algebraic relation between $q_c$ and the expansion parameters. Solving this relation for $p_c$ consistently in each
ordering then completes the specification of the corresponding asymptotic
basis.

Our starting point for the leading-order analysis is Eq.~\eqref{eq:mexp}. Substituting this solution into $c=m/r$ and writing
\begin{equation}
    \rho=\Bigl[\frac{3c}{4\pi\ec}\Bigr]^{-1/2}r
    \label{eq:rhodef}
\end{equation}
turns the relation between $c$ and $r$ into the cubic
\begin{equation}
    \rho^{3}-\rho+\tfrac{2}{3}\,\sigma_r=0 ,
    \label{eq:cubic}
\end{equation}
and similarly turns Eq.~\eqref{eq:mexp} into
\begin{align}
    \frac{p_c-p}{p_c}=c\biggl[\frac{\sigma_p}{q_c}
    &+\frac{(1+q_c)(1+3q_c)}{2q_c}\,\rho^{2}
    \nonumber\\
    &-\frac{2(1+q_c)}{3q_c}\,\frac{\sigma_r}{\rho}\biggr] ,
    \label{eq:pexact}
\end{align}
with $\sigma_p,\sigma_r$ of Eq.~\eqref{eq:sigmadef}. At this stage, no expansion in $\Sp$ or
$\Sr$ has been made. The leading-order solution depends on the two constants
only through the ratios $\sigma_p$ and $\sigma_r$, i.e., it is invariant under
\begin{align}
    &c\to\Lambda c,\qquad \Sp\to\Lambda\,\Sp,
    \nonumber\\
    &\Sr\to\Lambda^{3/2}\,\Sr,
    \qquad r\to\Lambda^{1/2}r ,
    \label{eq:rescaling}
\end{align}
for an arbitrary positive scaling parameter $\Lambda$, 
so that the three quantities entering on the same footing are
\begin{equation}
    c,\qquad \Sp,\qquad \Sr^{2/3} .
    \label{eq:threescales}
\end{equation}
The six orderings can now be classified in two steps. First, we examine
whether $\Sr$ is expanded before or after $c$. The radial solution is
determined by Eq.~\eqref{eq:cubic}, whose roots depend on the integration
constants only through $\sigma_r$; $\sigma_p$ does not enter at
this stage. It therefore suffices first to distinguish the regimes
$|\sigma_r|\ll 1$ and $|\sigma_r|\gg 1$. Second, after the segment solution
is obtained, we impose $p(C)=0$ and determine which of the three scales
$C$, $\Sp$, and $|\Sr|^{2/3}$ sets the leading behavior of the central
pressure.

In (A), (B), and (C), $\Sr$ is expanded
before
$c$, i.e., $|\sigma_r|\ll1$, and the root of Eq.~\eqref{eq:cubic} connected to
the regular solution is analytic,
\begin{equation}
    \rho=1-\frac{\sigma_r}{3}-\frac{\sigma_r^{2}}{6}+O(\sigma_r^{3}) .
    \label{eq:rhomatter}
\end{equation}
$\Sr$ enters through integer powers of $\sigma_r$, as in
Eqs.~\eqref{eq:rofc} and \eqref{eq:pofc}. The series \eqref{eq:rhomatter} has a finite radius of convergence. 
At $\sigma_r=3^{-1/2}$, the discriminant of Eq.~\eqref{eq:cubic} vanishes,
and the root \eqref{eq:rhomatter} merges with the second positive root at a
square-root branch point. For $\Sr>0$, the two positive roots exist only for
$\sigma_r\le 3^{-1/2}$. Thus, along the leading-order flow,
Eq.~\eqref{eq:rhomatter} exists only for $\sigma_r\le3^{-1/2}$, that is, only down to
$c_{*}=3^{1/3}\Sr^{2/3}$. 
Since Eq.~\eqref{eq:cubic} is itself a leading-order relation, this condition
specifies the domain of the leading-order solution needed to compare the
different orderings; it is not a bound on $\sigma_r$ for the full stellar
solution. The full expansion also has a finite radius of convergence in
$\sigma_r$, with $3^{-1/2}$ giving its leading-order estimate.

In (D), (E), and (F), $c$ is expanded before $\Sr$, i.e.,
$|\sigma_r|\gg1$. This regime supplies a leading-order
seed at small $c$ only when $\delta m<0$, and there
\begin{align}
    \rho&=\Bigl(\tfrac{2}{3}|\sigma_r|\Bigr)^{1/3}+O\bigl(|\sigma_r|^{-1/3}\bigr),
    \nonumber\\
    r&=\Bigl[\frac{3\,|\delta m|}{4\pi\ec}\Bigr]^{1/3}
    \Bigl[1+O\bigl(c/|\Sr|^{2/3}\bigr)\Bigr] .
    \label{eq:rhopoint}
\end{align}
In this regime, the $c^{1/2}$ scaling of Eq.~\eqref{eq:rofc} no longer
applies. Instead, $r$ approaches the radius for which the regular solution
of the segment encloses the mass $|\delta m|$. The dependence on $\Sr$ is
then governed by the single scale $|\Sr|^{2/3}$, with corrections organized
in powers of $c/|\Sr|^{2/3}$.

The surface condition is imposed only after the segment solution has been
obtained. Evaluating the pressure at the surface, where $c=C$, and setting
$p(C)=0$ gives an algebraic equation for $q_c=p_c/\epsilon_c$, which
determines the central pressure. Its asymptotic solution depends on the
ordering. Substituting the corresponding roots of Eq.~\eqref{eq:cubic} into
Eq.~\eqref{eq:pexact} gives four types of leading behaviors:
\begin{equation}
\begin{aligned}
    \text{(A)(B)}&:\quad q_c=q^{(1)}(\gamma)\,C\,
    \bigl[1+O(C,\sigma_p,\sigma_r)\bigr],\\
    \text{(C)}&:\quad q_c=\Sp\,\bigl[1+O\bigl(C/\Sp,\ \sigma_r\bigr)\bigr],\\
    \text{(D)(E)}&:\quad q_c=\Bigl[\bigl(\tfrac32\bigr)^{1/3}|\Sr|^{2/3}+\Sp\Bigr]
    \bigl[1+O(C)\bigr],\\
    \text{(F)}&:\quad q_c=\Sp\,
    \bigl[1+O\bigl(|\Sr|^{2/3}/\Sp,\ C/\Sp\bigr)\bigr] ,
\end{aligned}
\label{eq:qlocks}
\end{equation}
where $q^{(1)}(\gamma)$ denotes the leading  coefficient defined in Eq.~\eqref{eq:qcC}. 
Only in (A) and (B) does the surface condition give $q_c\propto C$ at
leading order, so that $q_c\to0$ as $C\to0$. The central pressure can
therefore be eliminated in favor of $C$, and the resulting stellar solution
can be organized as an expansion in $C$. In the other four orderings, the
leading value of $q_c$ is instead set by $\Sp$ and $\Sr$ and remains finite
as $C\to0$. Thus, $C$ no longer sets the leading scale of the solution.
This difference distinguishes the expansion bases relevant below.

Combining the local behavior of the segment solution with the scaling of
$q_c$ imposed by the surface condition completes the classification.
The six orderings do not give six distinct asymptotic structures:
(A) and (B) are equivalent at the level of the expansion basis, as are
(D) and (E), whereas (C) and (F) remain distinct. We therefore obtain four
inequivalent expansion bases,
\begin{align}
    &\text{I}\equiv\text{(A)(B)},\qquad
    \text{II}\equiv\text{(C)},
    \nonumber\\
    &\text{III}\equiv\text{(D)(E)},\qquad
    \text{IV}\equiv\text{(F)} .
    \label{eq:fourbases}
\end{align}
Within each paired case, exchanging the order of the two subleading
parameters changes only the ordering of terms within the corresponding
double expansion, not the set of monomials that defines the basis.

\subsubsection{Relation between expansion bases and the universal relation}
\label{sec:whichbasis}

For the universal relations, the relevant distinction among the four bases
is whether the compactness $C$ remains the leading expansion variable after
the surface condition is imposed.

In basis I, the central pressure is a function of the compactness,
Eq.~\eqref{eq:qlocks}, and every dimensionless observable $\cU_a$ becomes a series of the form
\begin{equation}
    \log\cU_a\bigl(C;\Sp,\Sr\bigr)=\delta_a\log C+O\bigl(1\bigr),
    \label{eq:logCtype}
\end{equation}
with the power $\delta_a$ fixed by the definitions of the observables alone
(details are given in Sec.~\ref{sec:Cexp}). Here $a$ is a label of the
observables. Eliminating $\log C$ between two observables leaves
$\log\cU_a=(\delta_a/\delta_b)\log\cU_b+\cdots$. This yields a straight line in the log--log plane
whose slope is fixed by the definitions, with corrections that are
parametrically small. For the log--log slope
\begin{equation}
    S\equiv\frac{d\log\cU_a}{d\log\cU_b}
    \label{eq:slopedef}
\end{equation}
this gives $S=\delta_a/\delta_b+O\bigl(C,\sigma_p,\sigma_r\bigr)$.

In bases II--IV that structure is absent. According to Eq.~\eqref{eq:qlocks} the central
pressure no longer varies with $C$ at leading order. It is instead fixed by the segment
constants. Consequently, every observable approaches an intercept determined by
$\Sp$ or $|\Sr|^{2/3}$ alone (Appendix~\ref{app:orderings}). The compactness
survives only inside the corrections, through ratios that are $O(1)$ precisely in
the domain of these bases, and the relation obtained after eliminating $C$ carries
no $\log\cU_b$ term at all. The slope \eqref{eq:slopedef} is therefore determined by the correction
terms and depends on $\gamma$, $\Sp$, and $\Sr$. Thus, unlike basis I,
bases II--IV do not generate a leading logarithmic behavior of universal relations.

The preceding comparison provides a criterion for selecting the relevant
basis: a universal relation requires an approximately constant leading slope
in the log--log plane. The known I--Love--Q and Love--$C$ relations exhibit
precisely this behavior over a wide range, with deviations described to a few
percent by low-order polynomials in $\log\cU$
\cite{Yagi:2013bca,Yagi:2013awa,Yagi:2016bkt}. Among the four bases, only
basis I has this structure. We therefore restrict the subsequent analysis to
basis I, corresponding to the regime in which the surface values
$\sigma_p(C)$ and $\sigma_r(C)$ of Eq.~\eqref{eq:sigmadef} are both small
compared with unity. This choice is not a microscopic restriction on the
equation of state, but the asymptotic regime consistent with the known
universal relations. Departures from this regime are discussed in
Sec.~\ref{sec:break}.

\subsubsection{The segment solution to all orders}

Above, we discussed the leading-order analysis: Eqs.~\eqref{eq:cubic} and \eqref{eq:pexact}
give all leading-order segment solutions, and
Sec.~\ref{sec:orderings} sorts them into the four bases
\eqref{eq:fourbases}. 
We now extend the construction to all orders.

Because only basis I can support a universal relation, we
construct the all-order solution here for that basis alone, in which $\Sp$ and
$\Sr$ are expanded before $c$. The corresponding construction in the other
regimes of the parameter space is given in
Appendix~\ref{app:orderings}.

Since $\Sp$ and $\Sr$ enter Eqs.~\eqref{eq:rofc} and \eqref{eq:pofc}
with $c^{-1}$ and $c^{-3/2}$, respectively, the operations listed
above change the exponents of the monomials only additively.
Consequently, the monomials generated by the iteration remain within the set
\begin{align}
\Sp^{\,n_1}\Sr^{\,n_2}c^{\,k-n_1-\frac32 n_2}
=
\sigma_p^{\,n_1}\sigma_r^{\,n_2}c^{\,k},
\qquad
k,n_1,n_2\in\mathbb{N}_0 .
\label{eq:lattice}
\end{align}
The same closure applies to the $q_c$ dependence of the coefficients.
Since the recursion involves only algebraic operations on rational functions
of $q_c$, the Laurent-polynomial structure already present in
Eq.~\eqref{eq:pofc} is preserved at all orders. The expansion coefficients of each monomial in Eq.~\eqref{eq:lattice} are therefore given by
\begin{align}
    \cO_{(k;n_1,n_2)}(\gamma,q_c)
    &=\sum_{m\ge0}\cO_{(k;n_1,n_2;m)}(\gamma)\,
    q_c^{\,m-M_{\cO}},
    \label{eq:qclaurent}
\end{align}
where $M_\cO$ are exponents summarized in Appendix~\ref{app:expansion}.
The leading term, $m=0$, is the Newtonian limit, and each additional power of
$q_c$ corresponds to one higher post-Newtonian order. The recursion that generates these coefficients,
and the explicit values of $M_\cO$ for the background variables, are given in
Appendix~\ref{app:expansion}.

Altogether, each component of Eq.~\eqref{eq:Oset} takes the form
\begin{align}
    \cO\bigl(c;\Sp,\Sr\bigr)\propto
    p_c^{\dim[\cO]}&c^{\Delta_{\cO}}
    \sum_{k\ge0}\Biggl[\sum_{n_1,n_2\ge0}
    \cO_{(k;n_1,n_2)}(\gamma,q_c)
    \nonumber\\
    &\times\sigma_p^{\,n_1}\,\sigma_r^{\,n_2}\Biggr]\,c^{\,k},
    \label{eq:cexp}
\end{align}
where $\Delta_{\cO}$ is the leading power of each variable, and $\dim[\cO]$ denotes the power of $p_c$ required by the dimension of \(\cO\). These powers are not
free. They are fixed by the behavior at stellar center
\eqref{eq:centralbg} and \eqref{eq:centralpert}, and are listed in
Appendix~\ref{app:expansion}. 
The central conditions are fully accounted for by these leading powers
in Eq.~\eqref{eq:cexp}.
The two constants $\Sp$ and
$\Sr$, however, remain free for a general segment. Only in the innermost
segment does $c\to0$ correspond to the physical center of the star. There,
the conditions $p\to p_c$ and $r\to0$ require
$\Sp=0=\Sr$ by Eqs.~\eqref{eq:rofc} and \eqref{eq:pofc}, in agreement with
Sec.~\ref{sec:layer}.
Accordingly, Eq.~\eqref{eq:cexp} gives the general solution in a segment
with the central conditions imposed and all remaining integration constants retained.

\subsection{Surface conditions and the \texorpdfstring{$C$}{C}-expansion}
\label{sec:Cexp}

To obtain the complete solution for the star, Eq.~\eqref{eq:cexp} must be
supplemented by the surface conditions of Sec.~\ref{sec:obs}: $p=0$,
$\nu(R)=\log(1-2C)$, and the matching conditions
\eqref{eq:Jdef}--\eqref{eq:H0match} for the three perturbation sectors. They are imposed at $c=C$, which by
Eq.~\eqref{eq:flowvar} is the same point as $r=R$. A similar analysis was already carried out at leading order in the previous section, and the
following analysis extends it to all orders.

The first condition, $p=0$, determines the central pressure. Consider first the
part of Eq.~\eqref{eq:cexp} with $\Sp=0=\Sr$. For the pressure, write
\begin{align}
p(c)=p_c[1-\sum_{k\ge0}p_k(q_c)c^{\,k+1}] ,
\end{align}
where the coefficients $p_k(q_c)$ are
generated by the recursion described in Appendix~\ref{app:expansion}. For example, Eq.~\eqref{eq:pofc} gives
\begin{equation}
    p_0(q_c)=\frac{1}{2q_c}+2+\frac{3q_c}{2}.
\end{equation} The surface condition $p(C)=0$ then yields
\begin{align}
    1=\sum_{k\ge0}p_k(q_c)\,C^{\,k+1} .
    \label{eq:surfcond}
\end{align}
Solving
Eq.~\eqref{eq:surfcond} for $q_c$ gives
\begin{equation}
    q_c(C)=q^{(1)}(\gamma)\,C+O(C^{2}),
    \label{eq:qcC}
\end{equation}
with a $\gamma$-dependent constant $q^{(1)}$. The successive Laurent orders in
$q_c$ supply the successive post-Newtonian corrections, which is the hierarchy
used in Sec.~\ref{sec:VC}.

With $\Sp,\Sr\ne0$ the same inversion applies, with $p_k$ replaced by the double
series in $\sigma_p,\sigma_r$ of Eq.~\eqref{eq:cexp}. One subtlety should be noted. Solving the surface condition gives $p_c$ as a series in $C$, $\Sp$ and
$\Sr$, and this series has to be built from the monomials \eqref{eq:lattice}
again. If it were not, substituting it back into Eq.~\eqref{eq:cexp} would
produce powers outside that set. This closure is indeed preserved since the leading term of
Eq.~\eqref{eq:surfcond} is linear in $q_c$, so the equation can be solved for
$q_c$ order by order, and each step only adds monomials already present
(Appendix~\ref{app:expansion}). The
result is
\begin{align}
    p_c\bigl(C;\Sp,\Sr\bigr)
    &=\bigl(K_0\,q^{(1)}C\bigr)^{\frac{1}{1-\gamma_0}}
    \nonumber\\
    &\quad\times\sum_{k,n_1,n_2\ge0}(p_c)_{(k;n_1,n_2)}(\gamma_0)\;
    \sigma_p^{\,n_1}\,\sigma_r^{\,n_2}\;C^{\,k},
    \label{eq:pcC}
\end{align}
with $\sigma_p,\sigma_r$ evaluated at the surface as in
Eq.~\eqref{eq:sigmadef}. The first coefficient is normalized as $(p_c)_{(0;0,0)}=1$. The power $1/(1-\gamma_0)$ follows from 
$q_c=p_c^{1-\gamma_0}/K_0$ and the leading-order relation
$q_c=q^{(1)}(\gamma_0)C+O(C^2)$. The condition $\nu(R)=\log(1-2C)$ and
the three exterior matchings \eqref{eq:Jdef}--\eqref{eq:H0match} then fix the constants \eqref{eq:signorm} as series of the
same form, and substituting
these back into Eq.~\eqref{eq:cexp} converts the $c$-expansion of every $\cO$
into an expansion in $C$.

Since the observables are algebraic combinations of the components of
Eq.~\eqref{eq:Oset}, they inherit the same series structure. 
With the dimensionless observables denoted by $\cU_a$ and their logarithms taken, the expansion takes the form
\begin{align}
    &\log\cU_a=\log\cU_a^{\rm N}(\gamma;\Sp,\Sr)+\delta_a\log C
    \nonumber\\
    &\qquad\qquad+\sum_{k\ge1}A_{a,k}(\gamma;\Sp,\Sr)\,C^{\,k},\label{eq:UCexp}
\end{align}
where
\begin{align}
    A_{a,k}=\sum_{n_1,n_2\ge0}A_{a,(k;n_1,n_2)}(\gamma)\,\Sp^{\,n_1}\Sr^{\,n_2}\,C^{-n_1-\frac32 n_2} .
\end{align}
Here $\log\cU_a^{\rm N}$ is the
Newtonian-order result after the logarithmic term
$\delta_a\log C$ is separated.
The powers $\delta_a$ are fixed by the definitions of the observables alone and
do not depend on the equation of state. For example, in the Newtonian limit $I\sim MR^2$ and
$\lambda_{\rm tid}\sim R^5$, so Eq.~\eqref{eq:observables} gives
\begin{equation}
    \Ib=i(\gamma)\,C^{-2},
    \qquad
    \lbt=\frac{2}{3}\frac{k_2^{\rm tid}}{C^5},
    \qquad
    i(\gamma)\equiv\frac{I}{MR^{2}},
    \label{eq:Inewt}
\end{equation}
and
\begin{equation}
    \delta_{\Ib}=-2,\qquad \delta_{\Qb}=-1,\qquad \delta_{\lbt}=-5 .
    \label{eq:deltas}
\end{equation}
The functions $i(\gamma)$ and
$k_2^{\rm N}(\gamma)\equiv\lim_{C\to0}k_2^{\rm tid}$ are given in
Sec.~\ref{sec:newtonlimit}.

\subsection{Expansion of the universal relations}
\label{sec:elim}

To obtain a relation directly between two observables, as in the I--Love--Q
relations, we eliminate $C$
from Eq.~\eqref{eq:UCexp}. For a pair $(\cU_a,\cU_b)$, we first rewrite
the expansion for $\cU_b$ in terms of $\cU_b^{1/\delta_b}$:
\begin{align}
    \cU_b^{1/\delta_b}&=(\cU_b^{\rm N})^{1/\delta_b}\,C\,
    \exp\Bigl[\tfrac{1}{\delta_b}\textstyle\sum_{k\ge1}A_{b,k}C^k\Bigr]
    \nonumber\\
    &=b_0C\bigl(1+b_1C+b_2C^2+\cdots\bigr),\qquad b_0\neq0 .
    \label{eq:xseries}
\end{align}
The right-hand side is a formal power series in $C$ with nonvanishing linear
coefficient, so by the Lagrange inversion theorem, $C$ can be expressed uniquely as a power series in
$\cU_b^{1/\delta_b}$. Substituting the resulting series for $C$ into Eq.~\eqref{eq:UCexp} for
$\cU_a$ and eliminating the corresponding $\log C$ term gives
\begin{align}
    \log\cU_a=\frac{\delta_a}{\delta_b}\log\cU_b
    +\sum_{k,n_1,n_2\ge0}&\cR^{\log}_{(k;n_1,n_2)}(\gamma)\;
    \Sp^{\,n_1}\,\Sr^{\,n_2}
    \nonumber\\
    &\times\cU_b^{\,\frac{k-n_1-\frac32 n_2}{\delta_b}} .
    \label{eq:master_series}
\end{align}
The exponent $k-n_1-\frac32n_2$ in the powers of $\cU_b$ is precisely
that of the monomials \eqref{eq:lattice}. The powers of $\Sp$ and $\Sr$ in the segment problem
directly determine the powers that appear in the relation between the observables. The
coefficient of the first term follows from Eq.~\eqref{eq:deltas},
\begin{equation}
    \frac{\delta_{\Ib}}{\delta_{\lbt}}=\frac{2}{5},\qquad
    \frac{\delta_{\Qb}}{\delta_{\lbt}}=\frac{1}{5},\qquad
    \frac{\delta_{\Qb}}{\delta_{\Ib}}=\frac{1}{2},
    \label{eq:slopes}
\end{equation}
independent of $\gamma$, $\Sp$, and $\Sr$.

As an explicit example, we take
$(\cU_a,\cU_b)=(\Ib,\lbt)$, for which
Eq.~\eqref{eq:master_series} becomes explicit: $\delta_b=-5$, so the powers of
$\cU_b$ are $\lbt^{-k/5}$. 
If the coefficients independent of $\Sp=0=\Sr$ are denoted as
\begin{equation}
    M_k(\gamma)\equiv\cR^{\log}_{(k;0,0)}(\gamma),
    \label{eq:Mkdef}
\end{equation}
Eq.~\eqref{eq:master_series} becomes
\begin{equation}
    \log\Ib=\frac{2}{5}\log\lbt
    +\sum_{k\ge0}M_k(\gamma)\,\lbt^{-k/5}
    +\bigl[\text{terms in }\Sp,\Sr\bigr] .
    \label{eq:Ilambda}
\end{equation}
The expansion variable is $\lbt^{-1/5}$ rather than $C$. This is a consequence of
the inversion \eqref{eq:xseries}.
In Eq.~\eqref{eq:master_series}, the equation-of-state dependence enters a universal
relation only through the three segment parameters: the coefficient functions
$\cR^{\log}_{(k;n_1,n_2)}(\gamma)$ and the constants $\Sp$, $\Sr$ that
multiply them. 
For the $\Ib$--$\lbt$ relation, these coefficients split into the $M_k(\gamma)$ of
Eq.~\eqref{eq:Ilambda} and the terms containing $\Sp,\Sr$.

We emphasize that, in this derivation, it is \textit{unnecessary} to single out a particular pair.
Equation~\eqref{eq:master_series} holds for \textit{any} two dimensionless quantities built from the surface and exterior data, with the slope $\delta_a/\delta_b$ read off from their definitions through Eq.~\eqref{eq:deltas}, and the three ratios \eqref{eq:slopes} cover all three pairs in the I--Love--Q trio.
The Love--$C$ relation can be obtained directly from Eq.~\eqref{eq:UCexp}, 
which already relates $\lbt$ and $C$.
In the rest of the paper we work out the $(\Ib,\lbt)$ pair explicitly, since it is the one for which the coefficients are needed in our analysis below.
The coefficients of the $(\Qb,\lbt)$ and $(\Qb,\Ib)$ relations are collected in Appendix~\ref{app:fits}.

\section{Parameters governing the universality}
\label{sec:which}

Equation~\eqref{eq:Ilambda} shows that, for each segment, all equation-of-state
dependence of the relation is encoded in three quantities: $\gamma$, $\Sp$,
and $\Sr$.
Their appearance in the relation, however, does not by itself
imply a violation of universality. The universality is a property of the
curve \eqref{eq:curve} and not of the point on it.
A parameter variation whose effect is a displacement along the curve is
indistinguishable from a change of the central pressure and leaves the relation
intact. 
What measures the violation of universality is therefore the transverse displacement between the curves~\cite{Hu:2026}. 
In this paper we evaluate this displacement at fixed $\lbt$
and refer to it as the ``transverse'' component. In this section
we define this decomposition. 
We also show that the leading effect of $\Sp$ is indistinguishable from a
change in the central pressure and amounts to a displacement along the curve,
and the leading transverse response is therefore controlled by $\gamma$ and $\Sr$.

\subsection{Tangential and transverse components}
\label{sec:decomp}

\begin{figure}[t]
\centering
\includegraphics[width=0.78\columnwidth]{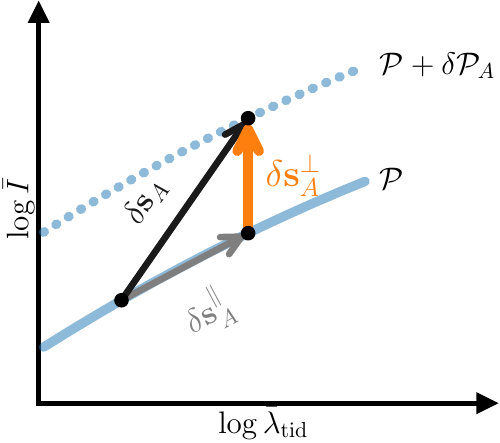}
\caption{Decomposition of a parameter variation. A star with parameters $\cP$
sits on the solid curve; changing $\cP_A$ at fixed central pressure carries it
by $\delta\mathbf{s}_A$ onto the dotted curve of $\cP+\delta\cP_A$, while
raising the central pressure moves it along the solid curve by
$\delta\mathbf{s}_A^{\parallel}$. The remainder $\delta\mathbf{s}_A^{\!\perp}$, taken at fixed
$\lbt$, is the transverse component \eqref{eq:perpdef}.}
\label{fig:transverse}
\end{figure}

The observables depend on the three segment parameters, i.e.,
\begin{equation}
    \cP\equiv\bigl(\gamma,\ \Sp,\ \Sr\bigr),
    \qquad \cP_A\ \ (A=1,2,3),
    \label{eq:Gmap}
\end{equation}
and on the central pressure, $p_c$. The former determine the curve and the latter selects a point on it.
We work in the observable plane, with $\log\lbt$ horizontal and $\log\Ib$
vertical.

At fixed $\cP$, raising the central pressure moves a star to the
neighboring member of the same sequence, that is, along the curve, by
\begin{equation}
    \delta\mathbf{s}_c=\Bigl(\frac{\partial\log\lbt}{\partial\log p_c},\
    \frac{\partial\log\Ib}{\partial\log p_c}\Bigr)\,\delta\log p_c ,
    \label{eq:dsc}
\end{equation}
whereas changing one of the segment parameters moves it to a different sequence
by
\begin{equation}
    \delta\mathbf{s}_A=\Bigl(\frac{\partial\log\lbt}{\partial\cP_A},\
    \frac{\partial\log\Ib}{\partial\cP_A}\Bigr)\,\delta\cP_A .
    \label{eq:dsA}
\end{equation}
A displacement along $\delta\mathbf{s}_c$ leaves the curve invariant, so $\cP_A$
moves the curve if and only if
$\delta\mathbf{s}_c\times\delta\mathbf{s}_A\neq0$.

We accordingly decompose
$\delta\mathbf{s}_A=\delta\mathbf{s}_A^{\parallel}+\delta\mathbf{s}_A^{\!\perp}$,
where $\delta\mathbf{s}_A^{\parallel}$ lies along $\delta\mathbf{s}_c$ and
$\delta\mathbf{s}_A^{\!\perp}$ has no $\lbt$ component. These two conditions
determine the decomposition and give
\begin{equation}
\begin{aligned}
    \delta\mathbf{s}_A^{\!\perp}
    &=\Bigl(0,\ \Bigl(\frac{\partial\log\Ib}{\partial\cP_A}\Bigr)^{\!\perp}
      \delta\cP_A\Bigr),
    \\[2pt]
    \Bigl(\frac{\partial\log\Ib}{\partial\cP_A}\Bigr)^{\!\perp}
    &\equiv\frac{\partial\log\Ib}{\partial\cP_A}
    -\frac{\partial\log\Ib}{\partial\log p_c}\;
     \frac{\ \partial\log\lbt/\partial\cP_A\ }
          {\ \partial\log\lbt/\partial\log p_c\ } .
\end{aligned}
\label{eq:perpdef}
\end{equation}
Thus $\delta\mathbf{s}_A^{\!\perp}$ is the separation of the two curves at fixed
$\lbt$ (Fig.~\ref{fig:transverse}). The transverse component \eqref{eq:perpdef} is
evaluated below for each of the three parameters.

\subsection{The regular solution and the tangential direction}
\label{sec:sigmap}
The parameter $\Sp$ is not a deformation of the star but a relabelling of the central pressure, and for $\Sr=0$ its transverse component vanishes identically. 
In fact, at $\Sr=0$ the integration constant $\delta m$ of Eq.~\eqref{eq:mexp} vanishes and the solution of the segment is the regular one. 
A regular solution of a given equation of state is fixed by its central pressure alone, which by Eq.~\eqref{eq:mexp} is $p_c-\delta p=p_c(1-\Sp/q_c)$. Every component of Eq.~\eqref{eq:Oset} therefore obeys
\begin{equation}
    \cO\bigl(c;\ p_c,\ \Sp,\ \Sr=0\bigr)
    =\cO_{\rm reg}\!\left(c;\ p_c\Bigl(1-\frac{\Sp}{q_c}\Bigr)\right),
    \label{eq:reparam}
\end{equation}
exactly. 
Differentiating at $\Sp=0$ yields
\begin{equation}
    \frac{\partial}{\partial\Sp}\bigl(\log\Ib,\log\lbt\bigr)\bigg|_{\Sr=0}
    =-\frac{1}{q_c}\frac{\partial}{\partial\log p_c}
      \bigl(\log\Ib,\log\lbt\bigr) ,
    \label{eq:dGdSp}
\end{equation}
so that $\Sp$ displaces the star by $-1/q_c$ times the tangent $\delta\mathbf{s}_c$. 
The variations of both observables are generated by the same derivative with respect to $p_c$, and Eq.~\eqref{eq:perpdef} gives
\begin{equation}
    \Bigl(\frac{\partial\log\Ib}{\partial\Sp}\Bigr)^{\!\perp}
    =-\frac{1}{q_c}\frac{\partial\log\Ib}{\partial\log p_c}
     +\frac{1}{q_c}\frac{\partial\log\Ib}{\partial\log p_c}=0 .
    \label{eq:sigmapperp}
\end{equation}

Since Eq.~\eqref{eq:reparam} holds for finite $\Sp$, the relation at
$\Sr=0$ is independent of $\Sp$ to all orders. In the master series
\eqref{eq:master_series}, this requires
$\cR^{\log}_{(k;n_1,0)}(\gamma)=0$ for all $n_1\ge1$. Hence every
$\Sp$-dependent transverse term also contains $\Sr$, and the transverse
contribution involving $\sigma_p$ starts at quadratic order,
$O(\sigma_p\sigma_r)$. The leading effect of varying $\Sp$ is therefore
a displacement along the curve. This justifies neglecting $\Sp$ at leading
order in Sec.~\ref{sec:layer}.

\subsection{Reduction to \texorpdfstring{$\gamma$}{gamma} and \texorpdfstring{$\Sr$}{Sr}}
\label{sec:remaining}

From Eqs.~\eqref{eq:perpdef} and \eqref{eq:sigmapperp}, under a simultaneous variation $\delta\cP=(\delta\gamma,\delta\Sp,\delta\Sr)$
the leading-order
transverse variation is
\begin{align}
    \bigl(\delta\log\Ib\bigr)^{\!\perp}
    &=\sum_{A}\Bigl(\frac{\partial\log\Ib}{\partial\cP_A}\Bigr)^{\!\perp}\delta\cP_A
    \nonumber\\
    &\simeq\Bigl(\frac{\partial\log\Ib}{\partial\gamma}\Bigr)^{\!\perp}\delta\gamma
    +\Bigl(\frac{\partial\log\Ib}{\partial\Sr}\Bigr)^{\!\perp}\delta\Sr .
    \label{eq:perplin}
\end{align}
Thus, the transverse displacement is controlled by $\gamma$ and $\Sr$.
The two coefficients are obtained directly from Eq.~\eqref{eq:master_series}.
At fixed $\lbt$, differentiation acts only on its coefficient functions,
since the term $\frac25\log\lbt$ carries no parameter dependence. Thus,
\begin{align}
    \Bigl(\frac{\partial\log\Ib}{\partial\gamma}\Bigr)^{\!\perp}
    &=\sum_{k\ge0}\frac{dM_k}{d\gamma}(\gamma)\;\lbt^{-k/5},
    \nonumber\\
    \Bigl(\frac{\partial\log\Ib}{\partial\Sr}\Bigr)^{\!\perp}_{\Sr=0}
    &=\sum_{k\ge0}\cR^{\log}_{(k;0,1)}(\gamma)\;
    \lbt^{\frac{3/2-k}{5}} .
    \label{eq:twoperp}
\end{align}
The two responses enter through distinct terms of
Eq.~\eqref{eq:master_series} and carry different powers of $\lbt$.
Substituting Eq.~\eqref{eq:twoperp} into Eq.~\eqref{eq:perplin} gives
\begin{align}
    \bigl(\delta\log\Ib\bigr)^{\!\perp}
    &=\sum_{k\ge0}\frac{dM_k}{d\gamma}(\gamma)\;
    \lbt^{-k/5}\;\delta\gamma
    \nonumber\\
    &\quad+\sum_{k\ge0}\cR^{\log}_{(k;0,1)}(\gamma)\;
    \lbt^{\frac{3/2-k}{5}}\;\delta\Sr .
    \label{eq:perpsum}
\end{align}
The leading-order transverse displacement therefore separates into
contributions from $\gamma$ and $\Sr$. A robust universal relation requires
both contributions to remain small, and we analyze them in the next two
sections.

\section{Result for the \texorpdfstring{$\gamma$}{gamma} dependence: reduction to a single polytrope}
\label{sec:gamma}

According to Eq.~\eqref{eq:perpsum}, the leading violation of the
universal relation is controlled by
$\gamma$ and $\Sr$, while the $\Sp$ contribution starts only at 
quadratic order. The integration constant $\Sr$ carries information inherited
from the inner segments through $\sigma_r$. This contribution decays 
as $C^{-3/2}$ with increasing compactness. 
We refer to this progressive suppression of the
interior dependence as \textit{memory loss}.
Once the inherited dependence is suppressed, the remaining equation-of-state
dependence within each polytrope segment is governed by the local index
$\gamma$ of that segment. This reduction can be written as
\begin{equation}
    \cU\bigl(C;\gamma,\Sp,\Sr\bigr)
    =\cU_{\rm sp}\bigl(C;\gamma\bigr)+O(\sigma_r,\ \sigma_p),
    \label{eq:reduction}
\end{equation}
where $\cU_{\rm sp}(C;\gamma)$ is obtained from the regular solution of the
segment, with the inherited integration constants removed. Extending this
regular solution with the same $\gamma$ throughout the star gives precisely
a single-polytrope star. 
Accordingly, the $\gamma$ dependence entering the universal relation is
determined by the corresponding single-polytrope equation of state.
In Eq.~\eqref{eq:Ilambda}, this dependence is encoded in the coefficients
$M_k(\gamma)$, which we determine by quadrature from a perturbative expansion
in $q_c$.
In this section
we first determine the range of $\gamma$ for which such stars exist and are
stable, and then evaluate the coefficients $M_k(\gamma)$. We set $\Sr=0$ throughout.

\subsection{Stability conditions for single-polytrope EoS}
\label{sec:range}

Equation~\eqref{eq:reduction} requires a stable single-polytrope sequence for
the corresponding value of $\gamma$. To determine the allowed range of
$\gamma$, it is sufficient to examine the low-compactness, Newtonian limit,
where the stability boundary can be obtained analytically. This Newtonian
argument is used only to identify the range of $\gamma$; the relativistic
stellar structure is retained in the subsequent calculation.

In the Newtonian limit, hydrostatic equilibrium and the definition of the
mass give
\begin{equation}
    p_c\sim\frac{\ec M}{R},\qquad M\sim\ec R^{3} ,
    \label{eq:balance}
\end{equation}
which, together with Eq.~\eqref{eq:eos}, yield
\begin{equation}
    R\sim K^{-\frac{1}{2\gamma}}\;\ec^{\frac{1-2\gamma}{2\gamma}},
    \qquad
    M\sim K^{-\frac{3}{2\gamma}}\;\ec^{\frac{3-4\gamma}{2\gamma}} .
    \label{eq:MRscaling}
\end{equation}
The equilibrium sequence is parametrized by the central density $\ec$.
Radial stability requires $dM/d\ec>0$ \cite{Shapiro:1983du}, so
Eq.~\eqref{eq:MRscaling} gives
\begin{equation}
    \frac{d\log M}{d\log\ec}
    =\frac{3-4\gamma}{2\gamma},
    \label{eq:stability}
\end{equation}
which is positive for $0<\gamma<3/4$. At $\gamma=3/4$, the mass becomes
independent of $\ec$, marking the marginally stable configuration and the
Chandrasekhar limit \cite{Chandrasekhar:1931ih}. The endpoint $\gamma=0$ is well defined separately: it corresponds to the
uniform-density star, although Eq.~\eqref{eq:stability} does not apply there~\cite{Schwarzschild:1916b}. We therefore restrict the
analysis to
\begin{equation}
    0\le\gamma<\frac34 .
    \label{eq:range}
\end{equation}

\subsection{Analytic expansion of the universal relation and its \texorpdfstring{$\gamma$}{gamma} dependence}
\label{sec:VC}

Having established the stable range of $\gamma$, we next quantify its
effect on the universal relation through the coefficients $M_k(\gamma)$.
 For the $(\Ib,\lbt)$ pair, the necessary equations are
\eqref{eq:TOV}, \eqref{eq:omegadiff} and \eqref{eq:H0diff}. For a single-polytrope equation of state, the constant $K$ drops out of
these equations, so the solution depends only on $\gamma$ and $q_c$
(Appendix~\ref{app:LE}). The quadrupole equations
\eqref{eq:k2diff} and \eqref{eq:h2diff}, which determine $\Qb$, are treated in the same
way (Appendix~\ref{app:fits}). 

According to Eq.~\eqref{eq:qcC}, $q_c=p_c/\ec$ is proportional to $C$ at leading order, as is $\lbt^{-1/5}$, so
Eq.~\eqref{eq:Ilambda} has a post-Newtonian hierarchy,
\begin{align}
    \log\Ib&=\underbrace{\tfrac{2}{5}\log\lbt+M_0(\gamma)}_{\text{Newtonian}}
    \;+\;\underbrace{M_1(\gamma)\,\lbt^{-1/5}}_{\text{1PN}}
    \nonumber\\
    &\quad+\;\underbrace{M_2(\gamma)\,\lbt^{-2/5}}_{\text{2PN}}\;+\;\cdots .
    \label{eq:PNhierarchy}
\end{align}
Each $M_k$ is thus fixed by the $k$-th post-Newtonian order. In the
following we therefore expand about the Newtonian limit.
Section~\ref{sec:newtonlimit} gives the closed-form solution in the Newtonian
limit and determines $M_0$. In Sec.~\ref{sec:PN}, expanding the equations in
powers of $q_c$ yields a hierarchy of linear equations, with each order
sourced by the lower-order solutions, from which the higher coefficients
$M_k$ are obtained recursively.
A related Newtonian analysis was presented in Ref.~\cite{Yip:2017}.

\subsubsection{Newtonian limit}
\label{sec:newtonlimit}

The problem is simplified in the Newtonian limit $q_c\to0$.
Following Refs.~\cite{Tooper2,Tooper1}, we introduce the new variables
\begin{align}
    &\theta\equiv\Bigl(\frac{p}{p_c}\Bigr)^{1-\gamma}=\frac{p/\epsilon}{q_c},
    \qquad
    \xi\equiv\frac{r}{a},
    \nonumber\\
    &a^{2}\equiv\frac{p_c}{4\pi(1-\gamma)\,\ec^{2}},
    \qquad
    \mu\equiv-\xi^{2}\frac{d\theta}{d\xi},
    \label{eq:LEvars}
\end{align}
so that $\epsilon/\ec=\theta^{\,n}$ with the polytrope index
$n\equiv\gamma/(1-\gamma)$ and $m=4\pi a^{3}\ec\mu$. Then, Eq.~\eqref{eq:TOV} reduces to the Lane--Emden equation of index $n$
\cite{Chandrasekhar:1939},
\begin{equation}
    \frac{d\mu}{d\xi}=\xi^{2}\theta^{\,n},
    \qquad
    \frac{d\theta}{d\xi}=-\frac{\mu}{\xi^{2}} ,
    \label{eq:LEnewt}
\end{equation}
integrated from $\theta=1$ at the center to the surface $\theta=0$ at
$\xi=\xi_1$. The constant $K$ has dropped out, and the Newtonian structure
depends on $\gamma$ alone.

The two perturbation sectors, \eqref{eq:omegadiff} and \eqref{eq:H0diff}, simplify in the same way, and likewise depend
on $\gamma$ alone. 
With the $\ell=2$
perturbation of the gravitational potential written as $\delta\Phi=H(r)Y_{2m}$ and using
$\delta\epsilon=-\epsilon\,(d\epsilon/dp)\,H$, which expresses hydrostatic balance of
the perturbed star, Poisson's equation gives
\begin{equation}
    G=\frac{dH}{d\xi},
    \qquad
    \frac{dG}{d\xi}=-\frac{2}{\xi}\,G+\frac{6}{\xi^{2}}\,H
    -n\,\theta^{\,n-1}H ,
    \label{eq:radau}
\end{equation}
with the regular central behavior $H\propto\xi^{2}$, $G\propto2\xi$. The surface
value of the tidal variable of Eq.~\eqref{eq:ydef} is then
\begin{equation}
    y=\frac{\xi_1G_1}{H_1}-3\,\frac{\epsilon(R)}{\bar\epsilon(R)},
    \qquad
    \frac{\epsilon(R)}{\bar\epsilon(R)}=\frac{\theta_1^{\,n}\,\xi_1^{3}}{3\mu_1},
    \label{eq:ynewt}
\end{equation}
where the second term accounts for the surface density jump and quantities with a subscript ``$1$'' are evaluated at the surface $\xi=\xi_1$.
It survives only for
$\gamma=0$, and the Love number follows from the $C\to0$ limit of
Eq.~\eqref{eq:k2tid},
\begin{equation}
    k_2^{\rm N}=\frac{1}{2}\,\frac{2-y}{3+y} .
    \label{eq:k2N}
\end{equation}
In the rotational sector $\omega_1$ is
uniform at this order and $I=\frac{8\pi}{3}\int\epsilon\,r^{4}dr$, so that
\begin{equation}
    i(\gamma)\equiv\frac{I}{MR^{2}}=\frac{2}{3}\,\frac{\mathcal J_1}{\mu_1\,\xi_1^{2}},
    \qquad
    \mathcal J\equiv\int_0^{\xi}\theta^{\,n}\,\xi'^{4}\,d\xi' .
    \label{eq:iN}
\end{equation}

The $k=0$ coefficient of Eq.~\eqref{eq:Ilambda} follows by inserting
Eq.~\eqref{eq:Inewt}:
\begin{equation}
    M_0(\gamma)=\log i(\gamma)-\frac{2}{5}\log\!\left[\frac{2}{3}k_2^{\rm N}(\gamma)\right] .
    \label{eq:M0}
\end{equation}
For $\gamma=0$ one finds $\xi_1=\sqrt6$ and $y=-1$, and
Eqs.~\eqref{eq:k2N}--\eqref{eq:iN} reproduce the uniform-density values
$k_2^{\rm N}=3/4$ and $i=2/5$.

\subsubsection{Perturbative expansion in \texorpdfstring{$q_c$}{qc}}
\label{sec:PN}

We use the variables \eqref{eq:LEvars} to organize an expansion about the Newtonian limit.
In these variables, Eqs.~\eqref{eq:TOV}, \eqref{eq:omegadiff}, and
\eqref{eq:H0diff} again become independent of $K$ and depend only on $\gamma$ and $q_c$. The relativistic corrections are therefore obtained from
Sec.~\ref{sec:newtonlimit} by restoring the terms that depend on $q_c$. 

We introduce the collective notation for the perturbation variables as 
\begin{align}
    Y\in\{\theta,\mu,w,V,H,G\},
\end{align}
where $w=\omega_1/\omega_c$ and $V= u^{-1}dw/d\xi$.
These variables can be expanded 
in $q_c$ at fixed
$\xi$ as
\begin{equation}
    Y(\xi)=\sum_{j\ge0}Y^{(j)}(\xi)\;q_c^{\,j},
    \qquad Y\in\{\theta,\mu,w,V,H,G\},
    \label{eq:PNexp}
\end{equation}
and matching powers of $q_c$ turns the system into a chain of ordinary
differential equations for $Y^{(j)}(\xi)$. Here, $Y^{(0)}(\xi)$ are the solutions of 
Eqs.~\eqref{eq:LEnewt} and \eqref{eq:radau}, and at each higher order the coefficient satisfies a linear equation sourced
by the lower-order solutions. The resulting equations are lengthy and are given in Appendix~\ref{app:LE}. They are integrated from the center to the surface in the same manner as in the Newtonian case.
Evaluating the solutions at the surface, denoted by the subscript ``$1$'',
yields the observables,
\begin{align}
    &C=\frac{u\,\mu_1}{\xi_1},
    \qquad
    u\equiv\frac{q_c}{1-\gamma},
    \nonumber\\
    &\Ib=\frac{(1-\gamma)^{2}\;\xi_1^{4}\,V_1}
         {q_c^{2}\;\mu_1^{3}\,\bigl(6\,w_1+2\,u\,\xi_1V_1\bigr)}.
    \label{eq:yhat}
\end{align}
The tidal deformability $\lbt$ follows from the surface value of
$y$ through Eq.~\eqref{eq:k2tid}.
According to Eq.~\eqref{eq:yhat}, the surface values carry the leading factors
$C\propto q_c$, $\Ib\propto q_c^{-2}$, $\lbt\propto q_c^{-5}$. Since the powers
\eqref{eq:deltas} are fixed by the definitions of the observables alone, we
remove the leading factors and use $q_c^{2}\Ib$ and $q_c^{5}\lbt$:
\begin{equation}
    \log\Ib-\frac25\log\lbt
    =\log\bigl(q_c^{2}\Ib\bigr)-\frac25\log\bigl(q_c^{5}\lbt\bigr) ,
    \label{eq:fdef}
\end{equation}
whose right-hand side consists only of quantities finite at $q_c=0$;
Eq.~\eqref{eq:fdef} equals the combination
$\log\cU_a-(\delta_a/\delta_b)\log\cU_b$ of Eq.~\eqref{eq:master_series}.
On the other hand, Eq.~\eqref{eq:xseries} gives
$\lbt^{-1/5}=b_0q_c+O(q_c^2)$. Inverting this relation at the level of the
polynomial coefficients to express $q_c$ in terms of $\lbt^{-1/5}$ and inserting
into Eq.~\eqref{eq:fdef} yields
\begin{equation}
    \log\Ib=\frac25\log\lbt
    +\sum_{k\ge0}M_k(\gamma)\;\lbt^{-k/5} .
    \label{eq:mainrelation}
\end{equation}
Each $M_k$ is given in closed form by series inversion, as an explicit combination of the $q_c$-series
coefficients of $q_c^{2}\Ib$ and $q_c^{5}\lbt$. The general formula and its first
few terms, of which Eq.~\eqref{eq:M0} is the $k=0$ case, are collected in
Appendix~\ref{app:Mk}.

\subsubsection{\texorpdfstring{$\gamma$}{gamma} dependence of the I--Love--Q relation}
\label{sec:gammadep}

\begin{figure}[t]
\centering
\includegraphics[width=0.9\columnwidth]{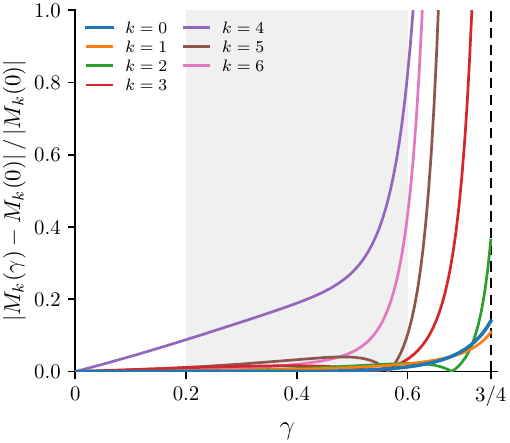}
\caption{Fractional change $|M_k(\gamma)-M_k(0)|/|M_k(0)|$ of the
coefficients of Eq.~\eqref{eq:mainrelation}, for $k\le6$. Shaded: the
realistic core band $\gamma\in[0.2,0.6]$ of Fig.~\ref{fig:gamma}; vertical
dashed line: the stability bound $\gamma=3/4$; the cusps touching zero are
crossings $M_k(\gamma)=M_k(0)$. All coefficients stay finite, but those with
$k\gtrsim3$ steepen sharply as $\gamma\to3/4$ and exceed the plotted range, reaching
$4.3$ for $k=3$ and $6.8\times10^{2}$ for $k=6$.}
\label{fig:Mk}
\end{figure}

\begin{figure}[t]
\centering
\includegraphics[width=\columnwidth]{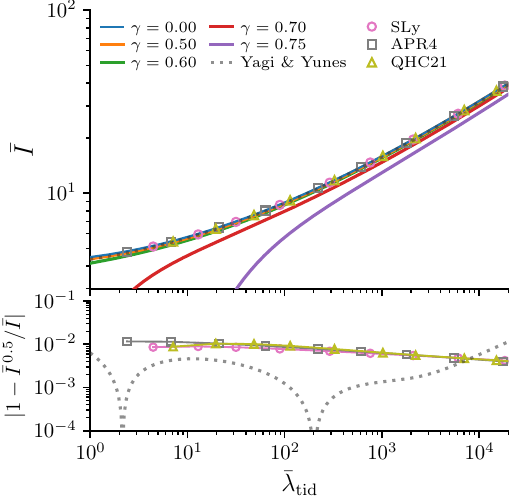}
\caption{$\Ib$ versus $\lbt$. Solid: Eq.~\eqref{eq:mainrelation} at
$\gamma=0,\ 0.50,\ 0.60,\ 0.70,\ 0.75$. Points: numerical sequences for SLy
\cite{Douchin:2001sv}, APR4 \cite{Akmal:1998cf} and QHC21 \cite{Kojo:2021wax},
each ending at its maximum mass. Dotted: the fitting formula of Yagi and
Yunes \cite{Yagi:2013awa}. Lower panel: fractional differences between these numerical results and $\Ib^{\,0.5}$. Here $\Ib^{\,0.5}$ denotes the exact result of
Eq.~\eqref{eq:mainrelation} at $\gamma=0.50$. The curves with $\gamma\le0.50$ are nearly indistinguishable.}
\label{fig:Ilambda}
\end{figure}

\begin{figure}[t]
\centering
\includegraphics[width=0.9\columnwidth]{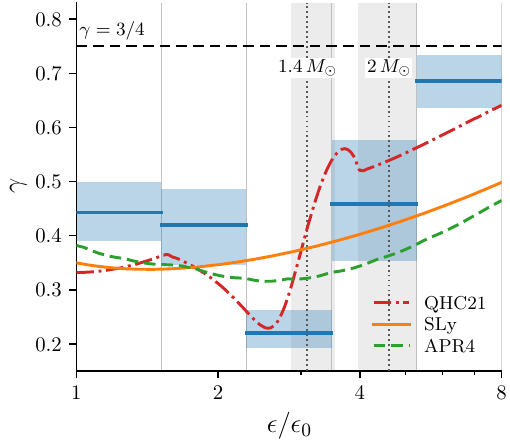}
\caption{Polytrope index $\gamma$ of Eq.~\eqref{eq:eos} as a function of energy density. Steps and bands: median and $68\,\%$ credible interval of the posterior of
Ref.~\cite{Fujimoto:2024cyv} on its five segments. Curves: SLy
\cite{Douchin:2001sv}, APR4 \cite{Akmal:1998cf} and QHC21 \cite{Kojo:2021wax}.
Dashed: the stability bound $\gamma=3/4$ of Eq.~\eqref{eq:stability}. Dotted
lines and gray strips: the central densities \eqref{eq:epsc} of $1.4$ and
$2\,M_\odot$ stars with their $68\,\%$ credible intervals.}
\label{fig:gamma}
\end{figure}

\begin{table*}[t]
\centering
\caption{Coefficients $M_k(\gamma)$ of the I--Love relation in the realistic
core band $\gamma\in[0.2,0.6]$.}
\label{tab:coeff}
\begin{tabular}{l|rrrrrrr}
\hline\hline
$\gamma$ & $M_0$ & $M_1$ & $M_2$ & $M_3$ & $M_4$ & $M_5$ & $M_6$ \\
\hline
0.20 & $-0.638834$ & 2.731671 & $-0.742927$ & 0.174472 & $-0.011346$ & $-0.035265$ & 0.018386\\
0.30 & $-0.638715$ & 2.726930 & $-0.741075$ & 0.175172 & $-0.011857$ & $-0.034927$ & 0.018281\\
0.40 & $-0.638961$ & 2.718963 & $-0.737652$ & 0.175416 & $-0.012401$ & $-0.034489$ & 0.018080\\
0.50 & $-0.640617$ & 2.704943 & $-0.732216$ & 0.174310 & $-0.013292$ & $-0.034246$ & 0.017449\\
0.60 & $-0.647328$ & 2.676668 & $-0.726907$ & 0.166665 & $-0.019019$ & $-0.039344$ & 0.010431\\
\hline\hline
\end{tabular}
\end{table*}

Carrying out the numerical integration of Sec.~\ref{sec:PN} through the sixth
post-Newtonian order across the range \eqref{eq:range} of $\gamma$, we obtain
the coefficients $M_k(\gamma)$ for $k\le6$, shown in
Fig.~\ref{fig:Mk}. Substituting these coefficients into
Eq.~\eqref{eq:mainrelation} gives the I--Love relations shown in
Fig.~\ref{fig:Ilambda} for representative values of $\gamma$. The figure
also compares them with numerical sequences for SLy, APR4, and QHC21 and
with the fitting formula of Ref.~\cite{Yagi:2013awa}. The series is truncated
at $k=6$; since each successive term is suppressed by an additional factor
of $\lbt^{-1/5}$, the truncation becomes relevant only toward the
low-$\lbt$ end of the curves.

Figure~\ref{fig:Mk} shows that the coefficients have markedly different
sensitivities to $\gamma$. At $\gamma=0.6$, $M_0$ and $M_1$ change by
$1.3\,\%$ and $2.2\,\%$, whereas $M_4$ changes by $82\,\%$. Toward the
stability bound $\gamma=3/4$, the change reaches $14\,\%$ for $M_0$ and
two to three orders of magnitude for $k\gtrsim4$, while all coefficients
remain finite. These large variations at higher order have little effect on
the relation 
because they enter at progressively higher orders in $\lbt^{-1/5}$. 
Consequently, the curves in Fig.~\ref{fig:Ilambda} agree
within $1.2\,\%$ up to $\gamma=0.50$ and separate by $3.0\,\%$, $11\,\%$,
and $35\,\%$ at $\gamma=0.60$, $0.70$, and $0.75$, respectively.

The reference point $\gamma=0$ in Fig.~\ref{fig:Mk} is the uniform-density
star, for which the I--Love relation is known in closed form
\cite{Chan:2014tva,Chan:2015iou}. Our $M_k(0)$ reproduce this result.

The fitting formula of Ref.~\cite{Yagi:2013awa}, also shown in
Fig.~\ref{fig:Ilambda}, is
\begin{align}
    \log\Ib\big|_{\rm YY}&=1.47+0.0817\,\ell+0.0149\,\ell^{2}
    +2.87\!\times\!10^{-4}\,\ell^{3}
    \nonumber\\
    &\quad-3.64\!\times\!10^{-5}\,\ell^{4},
    \qquad \ell\equiv\log\lbt. \label{eq:yyfit}
\end{align}
This fit agrees with
Eq.~\eqref{eq:mainrelation} to better than $1\,\%$. 
In that formula, both the expansion variable and the coefficients are fit
parameters, whereas in Eq.~\eqref{eq:mainrelation} the powers
$\lbt^{-k/5}$ are fixed by Eq.~\eqref{eq:deltas} and the coefficients
$M_k$ are determined by quadratures. Correspondingly, the residual of the
fit in Fig.~\ref{fig:Ilambda} is nonmonotonic, whereas that of
Eq.~\eqref{eq:mainrelation} decreases monotonically with increasing $\lbt$,
consistent with the expansion in powers of $\lbt^{-1/5}$.

\subsubsection{Realistic range of the polytrope index}
\label{sec:realisticgamma}

The value of $\gamma$ realized in realistic neutron stars must ultimately be constrained by observations. Current posterior ensembles of equations of state combine information from neutron-star observations, heavy-ion collisions, and theoretical constraints at low and asymptotically high density~\cite{Miller:2019nzo,Raaijmakers:2019dks,Essick:2019ldf,Annala:2019puf,Fujimoto:2021zas,Huth:2021bsp,Brandes:2022nxa,Annala:2023cwx,Legred:2024,Fujimoto:2024cyv}. 
In this work, we use the ensemble of Ref.~\cite{Fujimoto:2024cyv}, in which
the equation of state is represented by a polytrope interpolation within each of five logarithmically
spaced segments between $\epsilon_0=150\,\mathrm{MeV/fm^{3}}$ and
$8\epsilon_0$. The ensemble employs the SLy4 crust below $\epsilon_0$,
requires a maximum mass of at least $2.01\,M_\odot$, and is inferred by
neural networks from twenty observed sources.
For each of the $3\times10^{5}$ equations of state in the ensemble, we
determine the effective polytrope index
$\gamma=\Delta\log\epsilon/\Delta\log p$ of Eq.~\eqref{eq:eos} on each
segment using $dp/d\epsilon=c_s^2$.

Figure~\ref{fig:gamma} shows the density dependence of the effective
polytrope index inferred from the posterior ensemble. The median decreases
from $\gamma=0.443^{+0.057}_{-0.053}$ to
$0.220^{+0.042}_{-0.027}$ and then increases to
$0.686^{+0.047}_{-0.050}$ in the innermost segment. The corresponding
$68\,\%$ credible interval remains below the stability bound
$\gamma=3/4$ throughout. The tabulated equations of state exhibit a
similar range of variation. For SLy and APR4, $\gamma$ increases gradually
from $\simeq0.35$ to $\simeq0.5$, whereas QHC21 decreases to
$\simeq0.23$ near its speed-of-sound peak at
$2.7\epsilon_0$ and subsequently increases to $0.64$.

The density range probed by observed neutron stars can be estimated from
their central densities,
\begin{equation}
    \frac{\epsilon_c(1.4\,M_\odot)}{\epsilon_0}
    =3.09^{+0.45}_{-0.23},
    \qquad
    \frac{\epsilon_c(2.0\,M_\odot)}{\epsilon_0}
    =4.61^{+0.66}_{-0.65},
    \label{eq:epsc}
\end{equation}
where the uncertainties denote the $68\,\%$ credible intervals. These
ranges are indicated by the dotted lines and gray bands in
Fig.~\ref{fig:gamma}. For densities below $5\epsilon_0$, a range that contains the central densities relevant to the observed stars, the posterior has a median
$\gamma\le0.46$ and a $68\,\%$ upper bound $\gamma\le0.58$. The
tabulated equations of state occupy a comparable range. We therefore take
$\gamma\in[0.2,0.6]$ as the representative range for realistic neutron-star
matter and use it in Fig.~\ref{fig:Mk} and Table~\ref{tab:coeff}. Over this
range, the I--Love relations in Fig.~\ref{fig:Ilambda} remain nearly
coincident.

Only the innermost and least constrained segment of the posterior
\cite{Fujimoto:2024cyv} approaches the stability bound $\gamma=3/4$.
Stars whose central densities reach this region can therefore exhibit a
larger dependence of the I--Love relation on $\gamma$. Quantitatively, the
half-spread of $\log\Ib$ over $\gamma\in[0.2,0.6]$ is
$1.0$--$1.2\,\%$ for $\log\lbt=5.8$--$7.4$, compared with
$9$--$14\,\%$ over the full stable range $0\le\gamma<3/4$. Thus, the weak
$\gamma$ dependence of the relation for realistic equations of state is
associated with the range of polytrope indices sampled by observed stars,
which remains well separated from the stability boundary.

\section{Result for the \texorpdfstring{$\Sr$}{Sr} dependence: realistic stars and first-order phase transitions}
\label{sec:Sr}

By the decomposition \eqref{eq:perpsum} of Sec.~\ref{sec:which}, the curve of the universal relation is displaced along two directions only, $\gamma$ and $\Sr$. In Sec.~\ref{sec:gamma}, the $\gamma$ direction was analyzed and found that, over the realistic band of the index $\gamma$, the spread stays at the one-percent level. 
This section deals with the remaining $\Sr$ direction.
$\Sr$ is the integration constant transmitted outward by a segment, and its value is set by the structure of the interfaces. Therefore, we first identify the physics that generates a large $\Sr$, then evaluate how far a large $\Sr$ displaces the $\Ib$--$\lbt$ curve, and finally discuss the implications for realistic equations of state by combining these results with those of Sec.~\ref{sec:gamma}.

\subsection{Mechanisms that enhance \texorpdfstring{$\Sr$}{Sr}}
\label{sec:mech}

At fixed $\gamma$, the remaining transverse response is controlled by
$\Sr$ through $\sigma_r=\Sr/c^{3/2}$.
We first consider how interfaces can generate a large $\Sr$ and how the
resulting contribution is transmitted to the stellar surface.

A first-order phase transition described by the Maxwell construction
provides the simplest example that generates a large $\Sigma_r$. Consider a sharp interface at radius $r_t$,
where the energy density decreases outward from $\epsilon_-$ to
$\epsilon_+$ at the transition pressure $p_t$. We also define
$c_t\equiv m(r_t)/r_t$.
At leading order, Eq.~\eqref{eq:Srread} gives
\begin{equation}
    \sigma_r(c_t)=\frac32\,
    \sqrt{\frac{\epsilon_+}{\bar\epsilon(r_t)}}\,
    \frac{\bar\epsilon(r_t)-\epsilon_+}{\bar\epsilon(r_t)} .
    \label{eq:sigmaPT}
\end{equation}
For a transition occurring within the leading-order regime,
$\bar\epsilon(r_t)=\epsilon_-[1+O(c_t)]$, so that
$\bar\epsilon(r_t)-\epsilon_+=\Delta\epsilon+O(c_t\epsilon_-)$, with
$\Delta\epsilon\equiv\epsilon_--\epsilon_+$. Thus, a density discontinuity
comparable to the mean density inside the interface can generate
$\sigma_r(c_t)=O(1)$.

Outside the interface, $\Sr$ is constant, and Eq.~\eqref{eq:sigmadef}
relates its value at the transition to that at the stellar surface:
\begin{equation}
    \sigma_r(C)=\sigma_r(c_t)
    \left(\frac{c_t}{C}\right)^{3/2}.
    \label{eq:sigmaflow}
\end{equation}
A contribution generated at the transition is suppressed at
the surface when $c_t/C\ll1$, whereas no such suppression occurs when
$c_t/C=O(1)$. Thus, a strong Maxwell
transition can produce a surface value of order unity when the density
discontinuity is sufficiently large and $c_t$ is not small compared with
$C$. The expansion in $\sigma_r$ ceases to be controlled when
\begin{equation}
    |\sigma_r(C)|
    =\frac{|\Sr|}{C^{3/2}}\sim1 ,
    \label{eq:regimeboundary}
\end{equation}
corresponding to $C\lesssim|\Sr|^{2/3}$.

We briefly consider more general interfaces. For a Gibbs construction of a first-order phase transition, the density discontinuity is replaced by a finite
mixed-phase region in which $p$ and $\epsilon$ vary continuously. In this
case Eq.~\eqref{eq:sigmaPT} cannot be applied at a single interface. The
value of $\Sr$ transmitted to the outer solution must instead be determined
by integrating through the mixed phase. After the transition region has
been crossed, its effect on the subsequent segments is encoded in the
resulting values of $\Sp$ and $\Sr$. The Maxwell result is recovered only
in the corresponding sharp-transition limit.

The same distinction applies more generally to interfaces satisfying the
continuity condition \eqref{eq:eoscont}. Such an interface has no
density-jump contribution of the form \eqref{eq:sigmaPT}; the resulting
$\Sr$ is instead determined by the continuous evolution through the
adjacent segments.

\subsection{Breakdown of the universality at large \texorpdfstring{$\Sr$}{Sr}}
\label{sec:break}

\begin{figure}[t]
\centering
\includegraphics[width=\columnwidth]{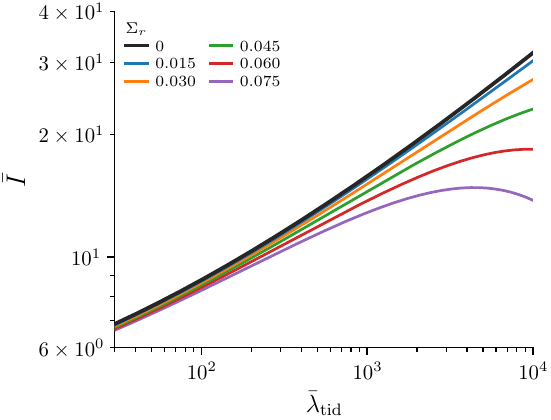}
\caption{Effect of the memory constant $\Sr$ on the $\Ib$--$\lbt$ relation,
from the analytic expression~\eqref{eq:normalresp} at $\gamma=0.5
$, for the indicated values of
$\Sr$. Here $\sigma_r(C)=\Sr/C^{3/2}$, and $C$ is obtained from $\lbt$ by the
inversion \eqref{eq:xseries}. The curves are displaced from the $\Sr=0$ curve
towards smaller compactness, where the same $\Sr$ produces a larger
$\sigma_r(C)$.}
\label{fig:sigmadev}
\end{figure}

\begin{figure}[t]
\centering
\includegraphics[width=\columnwidth]{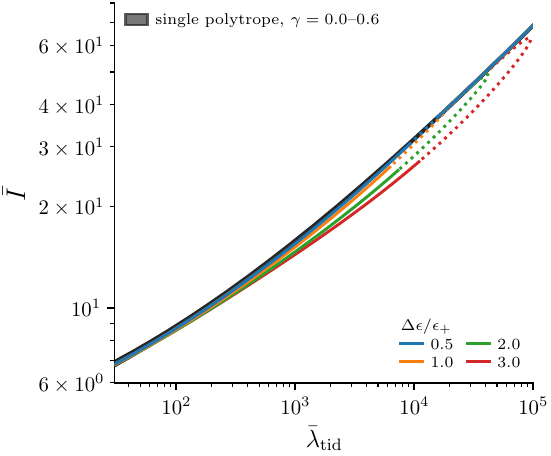}
\caption{
I--Love relations for the piecewise-polytrope EoS of
Eq.~\eqref{eq:two_segment_test}, consisting of two polytrope segments
separated by a sharp first-order phase transition. These curves 
are obtained by direct numerical integration. Both segments have
$\gamma=1/2$. We choose $K_0=1$ and $K_1=(1+j)K_0$, in which case
$j=\Delta\epsilon/\epsilon_+$. The curves correspond to
$\Delta\epsilon/\epsilon_+=0.5$, $1.0$, $2.0$, and $3.0$.
The transition pressure is fixed at $p_t=0.15\,p_c^{\rm ref}$, with
$p_c^{\rm ref}$ taken at $\bar\lambda_{\rm tid}=300$ on the
$\gamma=1/2$ single-polytrope sequence. The sequences are terminated at the
causal limit. Solid and dotted segments denote configurations with
$dM/dp_c>0$ and $dM/dp_c<0$, respectively. The gray band shows the
single-polytrope I--Love relations of Eq.~\eqref{eq:mainrelation} for
$0\leq\gamma\leq0.6$.
}
\label{fig:latent}
\end{figure}

Equation~\eqref{eq:regimeboundary} shows that a sufficiently strong
first-order transition can drive the surface value to
$|\sigma_r(C)|=O(1)$. 
We next quantify the resulting deviation from the universal relation.

According to Eq.~\eqref{eq:perpsum}, 
the transverse response to second order in
$\sigma_r$ can be written as
\begin{equation}
    \bigl(\delta\log\Ib\bigr)^{\!\perp}
    =\cA(C,\gamma)\,\sigma_r(C)+\cA_2(C,\gamma)\,\sigma_r(C)^{2}+O(\sigma_r^{3}) ,
    \label{eq:normalresp}
\end{equation}
where $\cA$ and $\cA_2$ denote the linear and quadratic transverse response
coefficients, respectively. Here, we also use $\lbt\sim C^{-5}$ to eliminate $\lbt$ in the coefficients. 
Small positive and negative mass offsets $\Sigma_r$ are introduced
into the regular single-polytrope solution, and the resulting transverse
displacements are evaluated relative to the unperturbed curve at fixed
$\lbt$.
The odd and even parts under $\Sr\to-\Sr$ isolate the linear and
quadratic terms, respectively. At $\gamma=0.5$, the resulting
coefficients are well described by
\begin{equation}
    \cA\simeq-0.017-0.17\,C,
    \qquad
    \cA_2\simeq-0.24 + 0.50\,C .
\end{equation}

Figure~\ref{fig:sigmadev} shows the corresponding response in the
$\Ib$--$\lbt$ plane. For fixed $\Sr$,
$|\sigma_r(C)|=|\Sr|/C^{3/2}$ increases toward smaller compactness.
Consequently, the deviation first becomes sizable at the low-$C$ end
of the sequence, corresponding to larger $\lbt$, rather than producing a
uniform displacement of the entire sequence. 

For the Maxwell transition considered in Sec.~\ref{sec:mech},
Eqs.~\eqref{eq:sigmaPT} and \eqref{eq:sigmaflow} give, within the
leading-order approximation,
\begin{equation}
    |\sigma_r(C)|
    \simeq
    \frac32
    \sqrt{\frac{\epsilon_+}{\bar\epsilon(r_t)}}\,
    \frac{\Delta\epsilon}{\bar\epsilon(r_t)}
    \left(\frac{c_t}{C}\right)^{3/2}.
\end{equation}
For $c_t\simeq C$, reaching the boundary
$|\sigma_r(C)|\sim1$ therefore requires
\begin{equation}
    \frac{\Delta\epsilon}{\bar\epsilon(r_t)}=O(1),
\end{equation}
provided that $\epsilon_+/\bar\epsilon(r_t)=O(1)$. Thus, in the absence of an additional enhancement from $c_t/C$, a sizable
deviation from the universal relation requires a density discontinuity of
the same order as the mean energy density enclosed by the transition.

To verify this behavior directly in stellar sequences, we consider an EoS
consisting of two polytrope segments separated by a sharp first-order phase
transition described by the Maxwell construction,
\begin{equation}
 \epsilon(p)=
 \begin{cases}
 K_0 p^\gamma, & p<p_t,\\
 K_1 p^\gamma, & p\geq p_t,
 \end{cases}
 \label{eq:two_segment_test}
\end{equation}
where $K_1\equiv(1+j)K_0$ and $j$ is an arbitrary number. 
In the numerical calculation, we use $\gamma=1/2$ and $K_0=1$. 
At $p=p_t$, we define
\begin{equation}
 \epsilon_+\equiv K_0p_t^\gamma,
 \qquad
 \epsilon_-\equiv K_1p_t^\gamma
 =(1+j)K_0p_t^\gamma ,
\end{equation}
so that
\begin{equation}
 \frac{\Delta\epsilon}{\epsilon_+}
 =\frac{\epsilon_--\epsilon_+}{\epsilon_+}=j .
 \label{eq:j_latent}
\end{equation}
Thus, varying $j$ changes the density discontinuity while leaving the
polytrope index unchanged.

Figure~\ref{fig:latent} shows the I--Love relations obtained by varying the
central pressure for several values of $\Delta\epsilon/\epsilon_+$. 
These curves are obtained by direct numerical integration of the full
stellar equations. The gray
single-polytrope band shows the range of the universal relation obtained by
varying the polytrope index over $0\leq\gamma\leq0.6$. For $p_c<p_t$, the
entire star is described by the low-density segment and the sequence follows
the single-polytrope relation. Once $p_c$ exceeds $p_t$, the density
discontinuity shifts the sequence away from the band, with a larger
$\Delta\epsilon/\epsilon_+$ producing a larger displacement. This behavior is
qualitatively consistent with the prediction in
Fig.~\ref{fig:sigmadev} and with the analysis of Sec.~\ref{sec:break}, where
a larger density discontinuity generates a larger $\Sr$ and hence a larger
transverse response. At still larger central pressures, the sequences turn back toward the band as the influence of the transition on the surface observables 
is reduced. This return is consistent with the memory-loss mechanism described
in Sec.~\ref{sec:gamma}.

The Maxwell transition discussed above gives $\Sr>0$. More generally,
Eq.~\eqref{eq:Srread} shows that the sign of $\Sr$ is determined by
$\bar\epsilon(r_t)-\epsilon_+$. A negative value, $\Sr<0$, corresponds
to a segment whose regular continuation toward the center contains more
mass than the actual interior. The behavior at large $|\Sr|$ then differs
qualitatively from the positive-$\Sr$ case.

For $\Sr>0$, the transverse displacement from the basis-I relation grows
with $\sigma_r$ until the expansion loses control near
Eq.~\eqref{eq:regimeboundary}. For $\Sr<0$, the physical branch can instead
be continued smoothly into the $|\sigma_r|\gg1$ regime. When $\Sp$ is
subleading, this regime is described by basis III of
Appendix~\ref{app:orderings}; basis IV applies when $\Sp$ is the dominant
segment constant. In these bases, Eq.~\eqref{eq:qlocks} no longer ties
$q_c$ to $C$ at leading order. Consequently, the definition-fixed
log--log slope characteristic of basis I is absent, and the relation acquires a different
expansion form. Large $|\Sr|$ therefore breaks the universality
for either sign, although the mechanism is different for $\Sr>0$ and
$\Sr<0$.

\subsection{Implications for realistic equations of state}
\label{sec:origin}

The size of the $\Sr$ contribution is determined by the response in
Eq.~\eqref{eq:normalresp} together with the surface value $\sigma_r(C)$.
As discussed in Sec.~\ref{sec:mech}, a large $\Sr$ can arise, for example,
from a strong first-order transition. Equation~\eqref{eq:sigmaflow} shows
that a contribution generated at small $c_t/C$ is suppressed before reaching
the stellar surface. A sizable displacement of the universal relation
therefore requires both a mechanism that generates a large $\Sr$ and
sufficiently weak suppression between the source region and the surface.

The realistic equations of state used in Sec.~\ref{sec:gamma} to test the universal relation
do not contain a strong first-order transition of the type considered in
Sec.~\ref{sec:mech}. This is consistent with the small breaking found for
these equations of state, since the mechanism identified above for producing
$|\sigma_r(C)|=O(1)$ is absent.

More generally, the breaking remains small when $\gamma$ stays away from
the stability bound and $|\sigma_r(C)|$ remains small. Significant
deviations may arise if either condition fails, for example if $\gamma$
approaches the stability bound or if a strong first-order transition
generates a large $\sigma_r(C)$. Conversely, within the present framework,
an observed deviation from the universal relation could indicate either
possibility. The latter is relevant, for example, to equations of state
that produce twin-star configurations \cite{Alford:2013aca}, for which the
analysis above sets the expected scale of the breaking. A quantitative
criterion for a significant deviation from the universal relation in terms
of the transition parameters $(\Delta\epsilon,p_t)$ requires a more complete
treatment.

These conclusions are not specific to the choice of $\Ib$ and $\lbt$.
For a general pair of dimensionless observables $(\cU_a,\cU_b)$, the
definition-fixed powers $\delta_a$ and $\delta_b$ follow from Eq.~\eqref{eq:deltas}, and
the coefficients of the corresponding relation are obtained from the
single-polytrope solutions in the same way as above. At leading order,
the $\gamma$ dependence is governed by
\begin{equation}
    M_0(\gamma)
    =\log\cU_a^{\rm N}(\gamma)
    -\frac{\delta_a}{\delta_b}
     \log\cU_b^{\rm N}(\gamma) ,
    \label{eq:recipe}
\end{equation}
provided that $\delta_b\ne0$. The higher-order coefficients $M_k(\gamma)$
then determine the corrections away from this limit. Thus, the degree of
universality for a given pair of observables is controlled by the
$\gamma$ dependence of the full set of coefficients $M_k$, with $M_0$
providing the leading contribution.

As a final remark, the numerical size of a deviation depends on
the normalization of the observables. Under
$\cU_a\to\cU_a^{\,n}$, $\delta_a$, $M_0$, and
$\bigl(\delta\log\cU_a\bigr)^{\!\perp}$ are all multiplied by $n$.
The normalization-independent measure is therefore
${\bigl(\delta\log\cU_a\bigr)^{\!\perp}}/{\delta_a}$.
All numerical values quoted in this paper use the normalization of
Eq.~\eqref{eq:observables}.

\section{Discussion}
\label{sec:disc}

The analysis in Secs.~\ref{sec:gamma}--\ref{sec:Sr} was carried out for a
specific class of static stellar models. In this section, we examine which
aspects of the construction remain robust when the underlying assumptions are
broadened and which require an extension of the present framework. We first
test the dependence on the representation of the equation of state, then
consider generalizations of the static stellar response, and finally classify
the mechanisms that lead to violations of universality in terms of the
additional degrees of freedom or input data involved. The resulting
classification is summarized in Table~\ref{tab:taxonomy}.

\subsection{Robustness against the piecewise-polytrope representation}
\label{sec:eosgen}

Piecewise polytropes provide a convenient
parametrization of realistic equations of state, which raises the question of whether the universality obtained above is
an artifact of the representation itself. Here we show that generalizing this representation
does not change the conclusions. 

To examine the dependence on the form of the equation of state within each
segment, we generalize Eq.~\eqref{eq:eos} to
\begin{equation}
    \epsilon(p)=\sum_{n\ge0}K_{(i,n)}\,p^{\,\gamma_i+n},
    \qquad p_i\le p<p_{i+1},
    \label{eq:eosgen}
\end{equation}
and apply the expansion of Sec.~\ref{sec:series}.

Substituting Eq.~\eqref{eq:eosgen} into Eq.~\eqref{eq:dpdc} gives
$p_c\sim(K_{(i,0)}C)^{1/(1-\gamma_i)}$. Each higher coefficient
$K_{(i,\ell)}$ therefore enters with an additional factor
$C^{\ell/(1-\gamma_i)}$, and Eq.~\eqref{eq:cexp} generalizes to
\begin{align}
    \cU
    &=K_{(i,0)}^{-\frac{\dim[\cU]}{1-\gamma_i}}\,
    C^{\delta_\cU}
    \sum_{k,\ell\ge0}\sum_{n_1,n_2\ge0}
    \cU_{(k,\ell;n_1,n_2)}
    \bigl(\gamma_i;\kappa_{i,\ell}\bigr)
    \nonumber\\
    &\qquad\times
    \sigma_p^{\,n_1}\sigma_r^{\,n_2}\,
    C^{\,k+\frac{\ell}{1-\gamma_i}},
    \label{eq:Ugenexp}
\end{align}
where
\begin{align}
    \kappa_{i,\ell}\equiv
    \left\{
    \frac{K_{(i,n)}}{K_{(i,0)}}\,
    K_{(i,0)}^{\,\frac{n}{1-\gamma_i}}
    \right\}_{1\le n\le\ell}.
    \label{eq:eosgenexp}
\end{align}
The $\ell=0$ sector reproduces Eq.~\eqref{eq:cexp}. All terms generated by
the higher coefficients in Eq.~\eqref{eq:eosgen} therefore enter with
positive additional powers of $C$, with stronger suppression at larger
$\ell$.

At leading order, only $\gamma_i$ and $K_{(i,0)}$ enter, whereas each
higher coefficient $K_{(i,\ell)}$ enters with an additional factor
$C^{\ell/(1-\gamma_i)}$. Over the range in Eq.~\eqref{eq:range},
$1/(1-\gamma_i)$ lies between $1$ and $4$. For example, at
$\gamma_i=0.5$, the first correction from Eq.~\eqref{eq:eosgen} enters at
$C^2$, one order beyond the first relativistic correction. Thus,
deviations from the single-power form affect only higher orders of the
compactness expansion and do not modify its leading structure.

The additional terms enter through the ratios
$\kappa_{i,\ell}$ as higher-order corrections to the coefficient
functions, while the equations for $(p,m)$ still contain the same two
integration constants. The counting of Sec.~\ref{sec:layer} and the memory
structure $(\Sp,\Sr)$ are therefore unchanged. After the overall scale set
by $K_{(i,0)}$ is factored out, the leading dependence on the equation of
state within a segment is still characterized by $\gamma_i$, as assumed in
Eq.~\eqref{eq:Gmap}. Thus, the main conclusions are unchanged under analytic
deformations of the piecewise-polytrope representation of the form
\eqref{eq:eosgen}, consistent with the numerical study of piecewise-polytrope
variations in Ref.~\cite{Benitez:2021}.

\subsection{Beyond barotropic
  perfect fluids and static response}
\label{sec:shape}
The mechanism of memory loss identified above suggests that the robustness of 
neutron-star universality depends on which information about the stellar interior 
survives in macroscopic observables. 
A natural question is therefore whether the 
same mechanism continues to suppress interior information when the framework is 
extended to more general stellar structure. 
Taking composition stratification, 
anisotropic stresses, elasticity, and magnetic fields as concrete examples 
\cite{Haskell:2013vha,Lau:2017,YagiYunes:2015aniso}, we examine 
whether the same integration constants continue to encode all
surviving interior information, or whether new independent information enters
the observables.

\subsubsection{Two speeds of sound and the static response}
\label{sec:twosounds}

A first possible source of additional interior information is composition
stratification. Equation~\eqref{eq:pertEOS} assumes that $\delta\epsilon$ is algebraically related to
$\delta p$. More generally, the stellar response can depend on additional thermodynamic
variables. In particular, in the presence of composition stratification,
two sound speeds are relevant: the equilibrium and adiabatic sound speeds,
\begin{equation}
    c_{\rm eq}^{2}=\frac{dp/dr}{d\epsilon/dr}, \qquad
    c_{\rm ad}^{2}\equiv\left(\frac{\partial p}{\partial\epsilon}\right)_{s,\,{\rm comp}}, \label{eq:ceq}
\end{equation}
which differ in general. Here $s$ denotes the entropy per particle and ``comp'' indicates fixed composition. With $\xi^{r}$ denoting the radial displacement, the
adiabatic condition reads
\begin{equation}
    \delta\epsilon=\frac{\delta p}{c_{\rm ad}^{2}}
    +\xi^{r}\,\frac{d\epsilon}{dr}\left[\frac{c_{\rm eq}^{2}}{c_{\rm ad}^{2}}-1\right] . \label{eq:deltaeps}
\end{equation}
The second term represents the density contrast relative to the surroundings, and produces
a restoring acceleration $-\cN^{2}\xi^{r}$, where, by Eq.~\eqref{eq:TOV},
\begin{equation}
    \cN^{2}=\frac{e^{\nu-\lambda}}{2}\,\frac{d\nu}{dr}\,\frac{1}{p+\epsilon}\,
    \frac{d\epsilon}{dr}\left[\frac{c_{\rm eq}^{2}}{c_{\rm ad}^{2}}-1\right] \label{eq:BV}
\end{equation}
is the squared relativistic Brunt--V\"ais\"al\"a frequency, whose sign determines the
Schwarzschild stability criterion~\cite{Ferrari:2002ut}. Because weak interactions are slow, composition stratification generally
leads to $c_{\rm ad}>c_{\rm eq}$ \cite{Reisenegger:2009}. 
Equation~\eqref{eq:deltaeps} then appears to introduce $\xi^{r}$ 
as an additional independent radial function, so that the closure assumed 
in Eq.~\eqref{eq:pertEOS} may seem to fail.

In the static limit, however, $\xi^{r}$ is not independent. The static
Euler equation, whose radial component is Eq.~\eqref{eq:TOV}, gives
\begin{equation}
    \nabla_i\,p=-\frac{p+\epsilon}{2}\,\nabla_i\nu
    \  , \label{eq:curl}
\end{equation}
from which we get $\nabla p\times\nabla\nu=0$.  
Thus, for a static perfect-fluid configuration in hydrostatic equilibrium, isobaric surfaces coincide with equipotential surfaces for any equation of state
\cite{Yagi:2013awa}. For the $\ell=2$ perturbation, this condition gives
$\delta\epsilon=\delta p/c_{\rm eq}^{2}$. Equation~\eqref{eq:deltaeps}
then determines $\xi^{r}$, so the system
\eqref{eq:dH0dc}--\eqref{eq:dbetadc} acquires no additional integration
constant. Thus, a barotropic equation of state is not required: the static response
depends only on the equilibrium derivative $d\epsilon/dp$, and the argument
of Sec.~\ref{sec:layer} applies equally to stratified stars, as confirmed in
Ref.~\cite{Penner:2011}. 
In this case, composition stratification therefore introduces no 
new independent information into the static observables.

Both staticity and the perfect-fluid assumption are essential to this
argument. At finite frequency, the density perturbation is governed by
$c_{\rm ad}$ rather than $c_{\rm eq}$, leading to a frequency-dependent
tidal response and to $g$ modes associated with $\cN^{2}$
\cite{Reisenegger:1992zm}. The additional stratification information then
enters the response dynamically.

Rotation provides another limitation. For rotating stars, the Euler 
equation has a first integral only for a barotropic fluid. 
Composition gradients can therefore provide additional radial information.
In newborn neutron stars, the composition is initially frozen and relaxes
on a timescale of order one second, during which the usual universal
relations can be modified \cite{Martinon:2014}. By contrast, universality
persists for sequences at fixed entropy and lepton fraction
\cite{Raduta:2020}. These examples illustrate the distinction introduced at the
beginning of this subsection: stratification does not spoil the static
memory counting when its additional variables are constrained, but
universality can be modified once an independent radial profile enters the
stellar response.

\subsubsection{Anisotropic stress}
\label{sec:thirdconstant}

We next relax the perfect-fluid assumption by allowing anisotropic stress,
\begin{align}
    &T_{\mu\nu}=(p+\epsilon)\,u_\mu u_\nu+p\,g_{\mu\nu}+\pi_{\mu\nu},
    \nonumber\\
    &\pi_{\mu\nu}u^{\nu}=0,\qquad \pi^{\mu}{}_{\mu}=0 .
    \label{eq:aniso}
\end{align}
The static Euler equation then contains the additional force density
$\nabla_{\mu}\pi^{\mu}{}_{i}$. Consequently, the integrability condition
\eqref{eq:curl} no longer follows, and isobaric and equipotential surfaces
need not coincide. The $\ell=2$ response can therefore depend on
anisotropic-stress variables in addition to the perfect-fluid equation of
state. This can also modify the approximate self-similarity of isodensity
surfaces associated with the I--Love--Q universality
\cite{Yagi:2014qua}, and pressure anisotropy is known to weaken the
universal relations \cite{YagiYunes:2015aniso}. The form of this
modification depends on how the anisotropic stress is determined.

A simple case is a quasi-local anisotropic-fluid model, in which the
radial--tangential pressure difference is specified in terms of the fluid
variables and the local compactness \cite{Horvat:2010xf}. Let $\alpha$
denote the parameter that controls the magnitude of this pressure
anisotropy for a given background configuration. The equations for
$(H_0,\beta)$ remain a closed second-order system, but their coefficients
depend on $\alpha$. The coefficients in the $C$-expansion therefore
acquire an additional dependence,
\begin{equation}
    \cO_{(k;n_1,n_2)}(\gamma,q_c)
    \longrightarrow
    \cO_{(k;n_1,n_2)}(\gamma,q_c;\alpha),
    \label{eq:aniso_coeff}
\end{equation}
and the coefficients of the universal relation become
$M_k(\gamma,\alpha)$. Their dependence on $\alpha$ describes the
anisotropy dependence found in Ref.~\cite{YagiYunes:2015aniso}.

Magnetic deformation provides a different example. In perturbative
treatments of relativistic magnetized stars, the magnetic field is first
determined on the spherical background from Maxwell's equations, and its
electromagnetic stress-energy then enters the $\ell=2$ stellar perturbation
equations as an inhomogeneous source
\cite{Konno:1999,Colaiuda:2007}. In the variables used here, the resulting
radial system can be written schematically as
\begin{align}
    \frac{dH_0}{dc}
    &=\frac{\beta\,r}{4\pi\epsilon r^2-c},
    \nonumber\\
    \frac{d\beta}{dc}
    &=g_1(c)H_0+g_2(c)\beta+s_B(c),
    \label{eq:magnetic-source}
\end{align}
where the terms proportional to $H_0$ and $\beta$ in
Eq.~\eqref{eq:dbetadc} are written as $g_1(c)H_0$ and $g_2(c)\beta$,
respectively, while $s_B(c)$ is the $\ell=2$ source generated by the
electromagnetic stress. For a fixed magnetic-field geometry with field
strength $B_0$, the electromagnetic stress is quadratic in the field, so
that
\begin{equation}
    s_B(c)=B_0^2\,\bar s_B(c),
    \label{eq:Bscale}
\end{equation}
where $\bar s_B(c)$ is fixed by the field geometry. The magnetic field
therefore modifies the $\ell=2$ solution through the particular solution
generated by $s_B$. 
Different field strengths and geometries can
consequently lead to different relations among the stellar observables
\cite{Haskell:2013vha}. This effect is particularly important for
$\Qb$: the magnetic quadrupole remains finite in the nonrotating limit,
whereas $\Qb$ is normalized by $\Omega^2$, so the magnetic contribution
is enhanced in slowly rotating stars.

Elastic matter differs from the preceding cases because the anisotropic
stress is itself part of the stellar response. The metric perturbations
are then coupled to elastic displacement and traction variables
\cite{Penner:2011,Gittins:2020,Lau:2017}. A minimal representation of one
radial stress channel is obtained by introducing
\begin{equation}
    \cS\equiv\frac{1}{\ec}\,\pi^{r}{}_{r}\Big|_{\ell=2},
    \label{eq:Sdefn}
\end{equation}
with the corresponding differential equations
\begin{subequations}
\begin{align}
    \frac{dH_0}{dc}
    &=\frac{\beta\,r}{4\pi\epsilon r^{2}-c},\\
    \frac{d\beta}{dc}
    &=g_1H_0+g_2\beta+g_3\cS,\\
    \frac{d\cS}{dc}
    &=g_4H_0+g_5\beta+g_6\cS .
\end{align}\label{eq:third}
\end{subequations}
Here $g_1$ and $g_2$ are the same coefficients as above, while
$g_3$--$g_6$ depend on the elastic properties of the material. The
explicit forms of these coefficients are fixed once the constitutive 
relation between the anisotropic stress and the elastic deformation 
is specified.
Equations~\eqref{eq:third} describes a single radial stress channel. General
elastic models contain coupled radial and tangential displacement and
traction variables. 
For such a problem, the differential equations \eqref{eq:third} take the
more complicated form~\cite{Penner:2011,Gittins:2020,Lau:2017}. 

If the constitutive relation is local and its material coefficients are
regular at the stellar center, $c=0$ remains a regular singular point of
the spherical-harmonic radial equations. If the couplings to $H_0$ and $\beta$ are subleading at the center, the
coefficient $g_6$ has the leading behavior
\begin{equation}
    g_6=\frac{\Delta_s}{c}+O(1),
    \qquad
    \cS\propto c^{\Delta_s}\bigl[1+O(c)\bigr],
    \label{eq:g6}
\end{equation}
where $\Delta_s$ is the index of the corresponding regular stress branch. Regularity requires $\Delta_s\ge0$: a regular $\ell=2$ stress
component need not vanish at the center. Ordinary linear elasticity, for example,
admits regular $\ell=2$ stress behaviors with $\Delta_s=0$ and $1$
\cite{Love:1927}. The allowed values of $\Delta_s$ are determined by the elastic
properties of the material. This branch supplements the perfect-fluid
behaviors $H_0\propto c$ and $H_0\propto c^{-3/2}$, with the latter
removed by central regularity.

The presence of an additional regular stress branch does not by itself
introduce an additional free parameter in the tidal deformation. When
the elastic stress is induced entirely by the external tidal field, the regular elastic solution is fixed by central regularity, by the
continuity conditions for the metric variables and the traction conditions
at fluid--elastic interfaces, together with the boundary conditions at the
stellar surface. The tidal deformability is therefore uniquely 
determined for elastic crusts~\cite{Penner:2011,Gittins:2020} 
and for stars that are solid to the
center \cite{Lau:2017}. The induced elastic stress can be nonzero, but its
strength is fixed together with the other perturbation variables.

The situation is different when the stellar configuration contains an
additional stress component that is not fixed by the field equations and
boundary conditions for the tidal deformation. Its radial profile then
constitutes additional input to the perturbation problem. If this input
is described by a single additional radial mode, the reduced system
\eqref{eq:third} contains one corresponding regular solution. The
coefficient of this solution is an additional integration constant,
which we denote by $\Ssh$:
\begin{equation}
    \cS=\Ssh\,c^{\Delta_s}\bigl[1+O(c)\bigr].
    \label{eq:Ssh}
\end{equation}
The constant $\Ssh$ therefore characterizes the additional stress input
within this one-parameter stress channel. A prescribed stress profile
with independent radial structure instead supplies a radial function
rather than a single additional constant.

Substituting Eq.~\eqref{eq:Ssh} into the enlarged radial system and
applying the same recursive series construction used to obtain
Eq.~\eqref{eq:lattice} adds the radial combination
$\Ssh c^{\Delta_s}$ to the expansion. The corresponding monomials are
therefore
\begin{equation}
    \Sp^{\,n_1}\Sr^{\,n_2}\Ssh^{\,n_3}
    c^{\,k-n_1-\frac32 n_2+\Delta_s n_3},
    \qquad
    \sigma_s\equiv\Ssh c^{\Delta_s}.
    \label{eq:lattice-s}
\end{equation}
Here $\Ssh$ is distinct from the hydrostatic memory $\Sr$, which is fixed
by the enclosed stellar structure. If the stress channel extends to the
stellar surface, its surface value is
\begin{equation}
    \sigma_s(C)=\Ssh C^{\Delta_s},
\end{equation}
which is independent of $C$ for $\Delta_s=0$. Equation~\eqref{eq:perpsum}
then schematically becomes
\begin{equation}
    \bigl(\delta\log\Ib\bigr)^{\!\perp}
    \simeq
    \frac{dM_0}{d\gamma}\,\delta\gamma
    +\cA(C,\gamma)\,\delta\sigma_r
    +\cA_s(C,\gamma)\,\delta\sigma_s ,
    \label{eq:blueprint-s}
\end{equation}
where $\cA_s$ is the transverse response coefficient associated with the additional stress channel.

These cases separate three effects of anisotropic stress on the universal
relations. A parameterized anisotropy changes the coefficients in the
$C$-expansion, a magnetic field adds an inhomogeneous source to the
$\ell=2$ equations, and an independently specified stress component can
supply either an additional integration constant or, for a general radial
profile, an additional function.
A quantitative determination of the resulting corrections for
specific elastic models requires solving the full coupled perturbation
equations and is left for future work.

\subsubsection{Radial screening of internal stress}
\label{sec:stressscreening}

The effect of anisotropic stress also depends on its location within the
star. When anisotropic stress is confined to the stellar interior and the
overlying layers are fluid, the exterior layers are again governed by the
perfect-fluid $\ell=2$ system
\eqref{eq:dH0dc}--\eqref{eq:dbetadc} for $(H_0,\beta)$. The effect of the
interior stress then enters the outer solution only through the values of
$(H_0,\beta)$ at the boundary with the outer fluid layer. For a fixed
tidal normalization, these boundary values determine the part of the
outer fluid solution inherited from the interior stress. 

Such screening by a fluid envelope is found explicitly in
Ref.~\cite{Lau:2017}: elasticity can substantially modify the I--Love
relation in fully solid stars, whereas the effect is strongly reduced
when a solid core is covered by a fluid envelope. This behavior is
similar to the mechanism identified in Sec.~\ref{sec:layer}: once the
elastic region ends, the effect of elasticity enters the outer
perfect-fluid solution only through the boundary values of $H_0$ and
$\beta$ at the fluid--elastic interface. In this sense, the fluid envelope
can reduce sensitivity to the elastic interior in much the same way that
an outer EoS segment reduces sensitivity to the deeper stellar structure.
A related distinction appears in Sec.~\ref{sec:thirdconstant}:
$\sigma_s(C)$ directly characterizes the surface stress only when the
stress channel extends to the stellar surface, whereas an internal
elastic region affects the outer perfect-fluid solution through the
matching at the fluid--elastic interface.

When anisotropic stress extends into the outer layers, the size of the
correction is instead controlled also by the elastic properties of the
matter and by the radial extent of the elastic region. Calculations for
realistic neutron-star crusts find small elastic corrections to the tidal
deformability \cite{Gittins:2020,Biswas:2019}, while models with a solid
region extending over a large fraction of the star can show larger
effects \cite{Pereira:2020}. Thus, the impact of anisotropic stress on the
universal relations depends not only on its strength, but also on where
in the star the stress is supported. The quantitative extent of such
screening is model dependent and requires solving the full coupled
perturbation equations, which is left for future work.

\subsection{Classification of the violations of universality}
\label{sec:taxonomy}

\begin{table*}[t]
\centering
\caption{Classification of the violations of universality.}
\label{tab:taxonomy}
\begingroup\footnotesize\setlength{\tabcolsep}{4pt}
\begin{tabular}{c|l|l}
\hline\hline
Type & Mechanism & Examples \\
\hline
I
& large $\Sr$
& strong first-order transitions; twin-star configurations \cite{Benitez:2021} \\
& & \\
IIa

& equations modified by 
& anisotropy \cite{Horvat:2010xf,YagiYunes:2015aniso};
  magnetic fields \cite{Konno:1999,Colaiuda:2007,Haskell:2013vha};\\
& additional parameters or sources
& 
  elasticity \cite{Lau:2017,Penner:2011,Gittins:2020};
  uniform rotation $\chi$ \cite{Chakrabarti:2013tca,Doneva:2013rha};\\
& &
  dynamical-tide frequency \cite{Pnigouras:2022};
  superfluid spin ratio \cite{Yeung:2021};\\
& &
  entropy and lepton fraction \cite{Raduta:2020};\\
& &
  modified-gravity couplings
  \cite{Yagi:2013awa,Doneva:2014faa,Gupta:2017vsl} \\
& & \\
IIb
& free constants remaining after 
& one-mode stress (Sec.~\ref{sec:thirdconstant});
  fermion--boson stars \cite{Lam:2025} \\
& equations and sources are fixed
& dark matter admixed neutron stars \cite{Cronin:2023};\\
& & \\
III
& additional radial function
& radial stress profile (Sec.~\ref{sec:thirdconstant});\\
& &
  magnetic-field configurations \cite{Haskell:2013vha};\\
& &
  differential rotation \cite{Bretz:2015rna};\\
& &
  frozen composition profiles and entropy gradients \cite{Martinon:2014} \\
\hline\hline
\end{tabular}
\endgroup
\end{table*}
According to Secs.~\ref{sec:eosgen} and \ref{sec:shape}, violations of
universality can be classified by how additional degrees of freedom
or external inputs enter the expansion. It is important to note that
relaxing the assumptions of the present analysis does not necessarily
introduce additional degrees of freedom
or inputs. The extension of the EoS
considered in Sec.~\ref{sec:eosgen} modifies higher-order terms in the
$C$-expansion without changing the memory structure $(\Sp,\Sr)$. Likewise,
composition stratification in a static perfect-fluid star introduces additional
thermodynamic variables but does not add independent radial information to the
static observables \cite{Penner:2011}.

Type I is the growth of the ordinary $\Sr$ channel analyzed in
Sec.~\ref{sec:Sr}. A sufficiently strong first-order transition can drive
$\sigma_r(C)$ outside the basis-I regime \cite{Benitez:2021}. The surface
observables can then retain a sizable dependence on the deep-interior EoS, and
the universal relation can break down.

Type IIa consists of cases in which additional physical parameters or sources
modify the coefficients, source terms, or coupled equations that determine the
stellar solution. Such parameters and sources characterize the physical model
or external conditions under which a stellar sequence is constructed.
Different choices can lead to different universal relations, so universality
should be examined for each specified choice of the relevant parameters and
sources. For example, in a quasi-local anisotropic-fluid model, the anisotropy
parameter $\alpha$ modifies the coefficient functions
\cite{Horvat:2010xf,YagiYunes:2015aniso}, while for a fixed magnetic-field
configuration the field strength $B_0$ controls the inhomogeneous magnetic
source \cite{Konno:1999,Colaiuda:2007,Haskell:2013vha}. In elastic models, the
constitutive relation and elastic properties modify the coupled $\ell=2$ tidal
equations and can change the universal relations
\cite{Lau:2017,Penner:2011,Gittins:2020}. Further examples are the
dimensionless spin $\chi$ in uniformly rotating stars
\cite{Chakrabarti:2013tca,Doneva:2013rha}, the frequency of the dynamical
tidal response \cite{Pnigouras:2022}, the superfluid spin ratio
\cite{Yeung:2021}, fixed entropy and lepton fraction \cite{Raduta:2020}, and
modified-gravity couplings
\cite{Yagi:2013awa,Doneva:2014faa,Gupta:2017vsl}.

Type IIb occurs when additional free constants are required to distinguish
physically different stellar solutions even after the differential equations,
their coefficients, and their source terms have been fixed. The one-parameter
stress channel of Sec.~\ref{sec:thirdconstant} provides one example, with
$\Ssh$ specifying the additional regular stress solution. Other examples arise
when additional matter or field components require independent central values
to specify a stellar solution. Dark matter admixed neutron stars described by
a two-fluid system can show substantial deviations from the standard
single-fluid I--Love--Q relations \cite{Cronin:2023}, while fermion--boson
stars provide an analogous example with an additional bosonic component
\cite{Lam:2025}.

Type III arises when the additional information is not a finite number of
parameters or free constants but an independent radial function. Examples
include a radial stress profile specified independently of the stellar
response, magnetic-field configurations \cite{Haskell:2013vha}, and
differential rotation \cite{Bretz:2015rna}. Frozen composition profiles and
entropy gradients in proto-neutron stars provide further examples
\cite{Martinon:2014}. In such cases, a finite number of variables
$(C,\gamma,\Sp,\Sr,\ldots)$ is insufficient to characterize the stellar
solution, and the present $C$-expansion cannot be applied directly. If the
additional function can be represented by a suitable finite number of modes,
an extension analogous to Type IIb may be possible. In the general case,
however, a new analysis retaining the functional dependence is required.

Table~\ref{tab:taxonomy} summarizes this classification. Some extensions can
be incorporated by adding a finite number of variables to the expansion.
Others require an additional function in the radial problem. The classification
therefore provides a criterion for determining how the present framework
should be extended when additional stellar physics is introduced. Applying
this criterion to concrete stellar models and determining the resulting
corrections to the universal relations are natural directions for future work.

\section{Conclusions}
\label{sec:concl}

We have derived asymptotic expansions of the I--Love--Q and Love--$C$
relations directly from the stellar structure equations and their boundary
conditions, for slowly rotating, tidally deformed stars described by
a piecewise-polytrope equation of state. Our aim was to determine from these
expansions how the EoS enters the relations and how strongly it affects them.
The equations allow several forms of the asymptotic expansion. We classified
them and found that only one class is compatible with the observed
universality. In that class, information about the EoS in the deeper stellar
layers reaches the observables only through three parameters of the outermost
segment: two integration constants and the polytrope index associated with
that segment. The values of the two integration constants are determined by
matching the stellar solution across the inner interfaces, and their
contributions to the surface observables decrease when the compactness
increases. The high-density EoS therefore affects the observables through
quantities transmitted across the interfaces, and the sensitivity to this
interior structure decreases as the compactness of the star increases.

At leading order, only two of the three parameters displace the relations
transversely. For $\Sr=0$, the first integration constant $\Sp$ is exactly a
reparametrization of the central pressure and moves a star along the curve
without changing the relation. Transverse terms involving $\Sp$ arise only
together with $\Sr$ and start at quadratic order. The leading transverse
dependence is therefore controlled by the polytrope index and $\Sr$.

We first examined how the universal relations depend on the polytrope index.
To isolate this dependence, we reduced the problem to a single polytrope and
computed the coefficients of the expansion by quadrature. Over the range
relevant to realistic cores, the relations shift by only about $1\,\%$,
whereas the shift reaches $9$--$14\,\%$ over the full stable range of the
polytrope index. The small shift reflects the fact that realistic cores remain
well below the stability bound on the polytrope index.

For $\Sr$, we identified the conditions under which it becomes large and how
its contribution is transmitted through the star to the surface observables.
A sharp first-order transition can generate a large $\Sr$, with the leading
contribution set by the density jump relative to the mean density enclosed by
the transition. Its contribution to the surface observables is suppressed for
transitions deep inside the star and is less suppressed for
transitions closer to the surface. We evaluated the resulting transverse
displacement of the relations to second order in $\sigma_r$ and found that it
grows toward smaller compactness. Within the leading-order estimate, we found
that a sizable breaking is associated with a density jump comparable to the
mean enclosed energy density.

We also examined how far the framework extends beyond the assumptions used in
the main analysis. This led us to classify violations of universality
according to how additional degrees of freedom or external inputs enter the
stellar problem. The classification in Sec.~\ref{sec:taxonomy} distinguishes
whether the departure arises from an enhancement of the existing memory
channel or from additional parameters, sources, free constants, or radial
functions. It therefore provides a criterion for determining how the present
framework should be extended when additional stellar physics is introduced.

More broadly, an important question is whether the reduction of sensitivity
to interior structure found here persists for more general stellar models and
other universal relations. The present analysis suggests that this question
can be addressed by identifying which degrees of freedom reach the surface
observables and how their contributions scale. Extending this viewpoint beyond
the present setting may provide a systematic way to determine when
universality emerges, when it breaks down, and which physical effects control
the departures from universality. Since our analysis is formulated directly
in terms of the differential equations and their boundary conditions, the
same strategy may also be applicable to relations among other stellar
observables and to other self-gravitating systems.


\begin{acknowledgments}
This work was supported by the Japan Society for
the Promotion of Science (JSPS) KAKENHI Grant No.
24KJ0985 (S.M.), Nos. 22H05118 and 25K07298 (S.K.), and
by the IIW, WINGS Program, The University of Tokyo (J.M.).
\end{acknowledgments}

\appendix

\section{Construction of the segment expansion}
\label{app:expansion}

This appendix collects the technical steps of Sec.~\ref{sec:general}: the
leading powers that implement the central conditions, the recursion that
generates the coefficients of Eq.~\eqref{eq:cexp} together with their Laurent
structure in $q_c$, the inversion of the surface condition, and the relation
between the orderings of the three small parameters and the four expansion
bases.

\subsection{Leading powers}
\label{app:Delta}

The exponents $\Delta_\cO$ of Eq.~\eqref{eq:cexp} are the powers with which
each variable vanishes or remains finite at the center. They follow
from the central behavior \eqref{eq:centralbg} and \eqref{eq:centralpert} once it
is rewritten in $c$. Since $4\pi\epsilon r^{2}\to3c$ at the center,
Eq.~\eqref{eq:rofc} at $\Sr=0$ gives $r\propto c^{1/2}$, and hence
\begin{equation}
    \cO\ \propto\ r^{\,2\Delta_\cO}\ \propto\ c^{\,\Delta_\cO},
    \qquad
    \frac{r}{4\pi\epsilon r^{2}-c}\ \propto\ c^{-1/2} .
    \label{eq:Deltarule}
\end{equation}
The first relation allows $\Delta_\cO$ to be read directly from
Eqs.~\eqref{eq:centralbg} and \eqref{eq:centralpert}: a variable approaching a
nonzero constant has $\Delta_\cO=0$, one vanishing as $r^{2}$ has
$\Delta_\cO=1$, and $r$ itself has $\tfrac12$. The second fixes the two
variables defined as radial derivatives, $\beta$ and $\phi$, whose defining
equations \eqref{eq:dH0dc} and \eqref{eq:dphidc} each carry one power of
$r/(4\pi\epsilon r^{2}-c)$ on the right-hand side. The resulting exponents are
listed in Table~\ref{tab:Delta}.

\begin{table}[h]
\centering
\caption{Leading powers $\Delta_\cO$ of Eq.~\eqref{eq:cexp}, obtained from
Eq.~\eqref{eq:Deltarule}.}
\label{tab:Delta}
\setlength{\tabcolsep}{4pt}
\begin{tabular}{l|cccccccc}
\hline\hline
$\cO$ & $p_c-p$ & $r$ & $\nb$ & $\omega_1$ & $\phi$ & $H_0$ & $\beta$
      & $K_2,\,h_2$ \\
\hline
$\Delta_\cO$ & $1$ & $\tfrac12$ & $0$ & $0$ & $\tfrac12$ & $1$ & $\tfrac12$
             & $1$ \\
\hline\hline
\end{tabular}
\end{table}

With these exponents, every series of the form \eqref{eq:cexp} is regular at the center, so the initial
conditions need not be imposed separately. The other exponent of
Eq.~\eqref{eq:cexp}, $\dim[\cO]$, is the dimension of $\cO$ in units of $p_c$, i.e. 
$1$ for a pressure and $-\tfrac12$ for r. For the variables whose overall
normalization is free,  this dimensional prefactor is conventional and can be absorbed into $\sigma_\cO$.

\subsection{Recursion and the Laurent structure in \texorpdfstring{$q_c$}{qc}}
\label{app:recursion}

Inserting Eq.~\eqref{eq:cexp} into Eqs.~\eqref{eq:dpdc}--\eqref{eq:mu2} and
matching the monomials \eqref{eq:lattice} produces, for each
$(k;n_1,n_2)$, one algebraic equation that is linear in the coefficient of that
monomial and otherwise involves only coefficients of strictly lower $k$. The
recursion is therefore triangular and is solved in a single pass. For
$\Sp=0=\Sr$ the pair $(p,r)$ and the metric function $\nb$ read
\begin{align}
    p(c)&=p_c\Bigl[1-\sum_{k\ge0}p_k(q_c)\,c^{\,k+1}\Bigr],
    \nonumber\\
    r(c)&=p_c^{-1/2}\sum_{k\ge0}r_k(q_c)\,c^{\,k+\frac12},
    \nonumber\\
    \nb(c)&=\sigma_{\nb}\sum_{k\ge0}N_{0,k}(q_c)\,c^{\,k},
    \label{eq:prNseries}
\end{align}
with
\begin{align}
    p_k(q_c)&=\sum_{n=0}^{2k+2}p_{(k,n)}\,q_c^{\,n-k-1},
    \nonumber\\
    r_k(q_c)&=\sum_{n=0}^{2k+2}r_{(k,n)}\,q_c^{\,n-k-\frac32},
    \nonumber\\
    N_{0,k}(q_c)&=\sum_{n=0}^{2k-1}N_{0,(k,n)}\,q_c^{\,n+1-k},
    \label{eq:laurentranges}
\end{align}
with the last relation applying for $k\ge1$. These are Eq.~\eqref{eq:qclaurent} made explicit, with
$M_p=k+1$, $M_r=k+\frac32$, and $M_{N_0}=k-1$. For example, the first coefficients are determined as
\begin{align}
    &p_0=\frac{1}{2q_c}+2+\frac{3q_c}{2},
    \qquad
    r_0=\sqrt{\frac{3q_c}{4\pi}},
    \nonumber\\
    &N_{0,0}=1,
    \qquad
    N_{0,1}=-\bigl(1+3q_c\bigr).
    \label{eq:lowcoef}
\end{align}
The first two reproduce Eqs.~\eqref{eq:rofc} and \eqref{eq:pofc} at $\Sr=0$ whereas the
last follows from Eq.~\eqref{eq:dNdc} with $4\pi\epsilon r^{2}\to3c$ and
$4\pi r^{2}p\to3q_c\,c$. Restoring $\Sp,\Sr\neq0$ replaces each
$\cO_k(q_c)$ by the double series
$\sum_{n_1,n_2}\cO_{(k;n_1,n_2)}(q_c)\,\sigma_p^{\,n_1}\sigma_r^{\,n_2}$ of
Eq.~\eqref{eq:cexp}, whose $(n_1,n_2)=(1,0)$ and $(0,1)$ members at $k=0$ are the
terms displayed in Eqs.~\eqref{eq:rofc} and \eqref{eq:pofc}.

The remaining sectors follow without further input. Their equations are linear in
the perturbation variables with coefficients built from $p$, $r$, $\epsilon(p)$,
so the same recursion determines $H_0,\beta,\omega_1,\phi$ and then, through the
sources of Eqs.~\eqref{eq:dk2dc}--\eqref{eq:mu2}, the four functions
$K_2^{S},h_2^{S},K_2^{H},h_2^{H}$.

The recursion has been carried out explicitly for the background pair through
$k\le2$ at first order in $(\Sp,\Sr)$, using the variables $\hat p=p/p_c$ and
$\mathfrak{r}=4\pi\ec r^{2}/(3c)$, in which the system contains no fractional powers.
Once $\Sp$ and $\Sr$ are fixed by the normalizations \eqref{eq:Sdef},
every coefficient is determined, and the resulting values agree with the
independent relation: $\mathfrak{r}=1-\frac{2}{3}\sigma_r+\cdots$, the square of
Eq.~\eqref{eq:rhomatter}. This confirms both the
set of monomials \eqref{eq:lattice} and the two-constants-per-segment counting at the
level of the full nonlinear equations.

\subsection{Inversion of the surface condition}
\label{app:inversion}

We write the surface condition $p(C)=0$ of Sec.~\ref{sec:Cexp} as
\begin{align}
    F\bigl(q_c;\,C,\sigma_p,\sigma_r\bigr)&\equiv
    1-\sum_{k,n_1,n_2\ge0}p_{(k;n_1,n_2)}(q_c)
    \nonumber\\
    &\qquad\times
    \sigma_p^{\,n_1}\sigma_r^{\,n_2}\,C^{\,k+1}
    =0 ,
    \label{eq:Fdef}
\end{align}
and substitute $q_c=\hat q\,C$. 
According to Eq.~\eqref{eq:laurentranges}, the most negative
power of $p_{(k;0,0)}$ is $q_c^{-(k+1)}$, so at $\sigma_p=0=\sigma_r$ the terms
that survive $C\to0$ at fixed $\hat q$ are exactly the leading Laurent
coefficients, and Eq.~\eqref{eq:Fdef} reduces to
\begin{equation}
    \Phi(\hat q;\gamma)\equiv1-\sum_{k\ge0}p_{(k,0)}\,\hat q^{\,-(k+1)}=0 .
    \label{eq:Phidef}
\end{equation}
Equation \eqref{eq:Phidef} is the leading, Newtonian, form of the surface
condition, and its root is the constant $q^{(1)}(\gamma)$ of
Eq.~\eqref{eq:qcC}. That root is simple, $\partial\Phi/\partial\hat q\neq0$,
which is the nondegeneracy invoked in Sec.~\ref{sec:Cexp}.

The implicit function theorem can therefore be applied to
Eq.~\eqref{eq:Fdef}: the solution $\hat q(C,\sigma_p,\sigma_r)$ is a formal power
series in $C$, $\sigma_p$, and $\sigma_r$, with coefficients that are Laurent in
$q^{(1)}$. Since $\sigma_p$ and $\sigma_r$ are themselves among the monomials
\eqref{eq:lattice}, so is the result. It is Eq.~\eqref{eq:pcC} upon using
$q_c=p_c^{1-\gamma_0}/K_0$. This is what allows $p_c(C)$ to be substituted
back into Eq.~\eqref{eq:cexp} without leaving the set \eqref{eq:lattice}, and
hence what makes the $C$-expansion \eqref{eq:UCexp} well defined.

\subsection{The four expansion bases}
\label{app:orderings}

This appendix gives the all-order form of the four bases identified in
Sec.~\ref{sec:orderings}, each obtained by running the construction of
Sec.~\ref{sec:cexp} from the corresponding leading-order solution of
Eqs.~\eqref{eq:cubic} and \eqref{eq:pexact}.

\medskip\noindent
\textbf{Basis I.} This is the basis of orderings (A) and (B), in which both
$|\sigma_p|$ and $|\sigma_r|$ are small. The seeds are
Eqs.~\eqref{eq:rofc} and \eqref{eq:pofc}, and the construction in the main text
gives the monomials \eqref{eq:lattice}, i.e., at the surface,
\begin{equation}
    \text{I}:\quad
    \cU=C^{\delta_\cU}\sum_{k\ge0}
    \left[\sum_{n_1,n_2\ge0}\cU_{(k;n_1,n_2)}(\gamma)\,
    \sigma_p^{\,n_1}\sigma_r^{\,n_2}\right]C^{\,k} .
    \label{eq:regimeI}
\end{equation}
The scaling \eqref{eq:rescaling} of the leading-order solution survives to all
orders: substituting Eq.~\eqref{eq:sigmadef} into Eq.~\eqref{eq:cexp} and
regrouping in powers of $c$,
\begin{align}
    \cO\;\propto\;p_c^{\dim[\cO]}
    \sum_{k,n_1,n_2\ge0}&\cO_{(k;n_1,n_2)}(\gamma)
    \nonumber\\
    &\times c^{\,\Delta_\cO+k-n_1-\frac32 n_2}\;\Sp^{\,n_1}\;\Sr^{\,n_2} ,
    \label{eq:cexp2}
\end{align}
so that the order-$k$ part $F_k$ obeys
\begin{equation}
    F_k\bigl(\Lambda c,\ \Lambda\Sp,\ \Lambda^{3/2}\Sr\bigr)
    =\Lambda^{\Delta_\cO+k}\,F_k\bigl(c,\Sp,\Sr\bigr) .
    \label{eq:homogeneity}
\end{equation}
Each $F_k$ is homogeneous of degree $\Delta_\cO+k$ in the three scales
\eqref{eq:threescales}. Homogeneity fixes which combined powers may appear; it
does not by itself decide the basis, because the function of the two ratios
multiplying the leading scale is analytic only within the regime in which the
series was constructed --- this is what distinguishes bases II--IV below from
mere regroupings of Eq.~\eqref{eq:regimeI}.

Eliminating $C$ from Eq.~\eqref{eq:regimeI} is the computation of
Sec.~\ref{sec:elim}: it returns the master relation \eqref{eq:master_series},
the slope $\delta_a/\delta_b$ of Eq.~\eqref{eq:slopes} together with corrections
that are ordered by the same monomials and are parametrically small throughout the
domain of the basis. The three cases below show what the same
elimination gives in the other regimes.

\medskip\noindent
\textbf{Basis II.} 
For the ordering (C), in which $|\sigma_r|\ll1\ll\sigma_p$, the
seed is still the root \eqref{eq:rhomatter} that is analytic at
$\sigma_r=0$, so $\Sr$ enters with integer powers of $\sigma_r$. The
surface condition \eqref{eq:qlocks} fixes $q_c=\Sp[1+\cdots]$. Since $q_c>0$, the physical branch considered here has $\Sp>0$, as does basis IV below. Running the construction with these seeds gives, at the
surface,
\begin{align}
    \text{II}:\quad
    \cU(C)&=\Sp\Bigl[\cO_{(0;0,0)}(\gamma)
    +\!\!\!\sum_{\substack{n_1\in\mathbb{Z},\ n_2,k\in\mathbb{N}_0\\ |n_1|+k>0}}
    \!\!\!\cO_{(k;n_1,n_2)}(\gamma)
    \nonumber\\
    &\qquad\times
    \Sp^{\,n_1}\,\Sr^{\,n_2}\,C^{\,k+|n_1|-\frac32 n_2}\Bigr] .
    \label{eq:regimeII}
\end{align}
Negative $n_1$ produces the ratios $(C/\Sp)^{|n_1|}$ in which the series is
ordered. The leading term carries no $C$: the surface condition has fixed
the central pressure in terms of $\Sp$, and every observable tends to the
$C$-independent limit $\cO_{(0;0,0)}\Sp$ as $C/\Sp\to0$.

There is therefore no power of $C$ to eliminate, and the inversion is performed
around that limit. The shifted variable
\begin{equation}
    \widetilde{\cU}_b\equiv\frac{\cU_b}{\Sp}-\cO^{\cU_b}_{(0;0,0)}
    \label{eq:shiftedII}
\end{equation}
vanishes linearly in $C$, so Lagrange inversion applies to it in place of
Eq.~\eqref{eq:xseries} and returns
$C=\widetilde{\cU}_b\bigl[C_{(0;0,0)}(\gamma)+\cdots\bigr]$. Substituting into
$\cU_a$ gives
\begin{align}
    \log\cU_a&=\log\Sp+\log\cM_{(0;0,0)}(\gamma)
    \nonumber\\
    &\quad+\!\!\!\sum_{\substack{n_1\in\mathbb{Z},\ n_2,k\in\mathbb{N}_0\\ |n_1|+k>0}}
    \!\!\!\cM^{\log}_{(k;n_1,n_2)}(\gamma)\,
    \Sp^{\,n_1}\,\Sr^{\,n_2}
    \nonumber\\
    &\qquad\times
    \widetilde{\cU}_b^{\;k+|n_1|-\frac32 n_2} ,
    \label{eq:relII}
\end{align}
where the series is ordered by $\widetilde{\cU}_b$ in the domain
$\Sr\,\widetilde{\cU}_b^{-3/2}\ll1$. No term proportional to $\log\cU_b$
survives.

\medskip\noindent
\textbf{Basis III.} 
For orderings (D) and (E), in which $|\sigma_r|\gg1$ with
$\Sp\lesssim|\Sr|^{2/3}$, the seed is the root \eqref{eq:rhopoint} together
with the surface condition of Eq.~\eqref{eq:qlocks}. At the next-to-leading order, the solution is given by
\begin{align}
    r(c)&=r_0+\frac{r_0^{2}}{3\,|\delta m|}\,c+O(c^{3}),
    \qquad
    r_0=\Bigl[\frac{3\,|\delta m|}{4\pi\ec}\Bigr]^{1/3},
    \nonumber\\
    \frac{p_c-p}{p_c}
    &=\frac{\Sp}{q_c}
    +\Bigl(\tfrac32\Bigr)^{1/3}\frac{(1+q_c)^{2}}{q_c}\,|\Sr|^{2/3}
    +\bigl(1+q_c\bigr)\,c
    \nonumber\\
    &\quad+\frac{2^{1/3}3^{2/3}(1+q_c)^{2}}{12\,q_c}\,\frac{c^{2}}{|\Sr|^{2/3}}
    +O\bigl(c^{3},\ c\,\Sp\bigr) .
    \label{eq:DEseed}
\end{align}
The generated monomials are closed on
\begin{align}
    \text{III}:\quad
    &\Sp^{\,n_1}\,|\Sr|^{\frac23 n_2}\,C^{\,k-n_1-n_2},
    \nonumber\\
    &n_1,k\in\mathbb{N}_0,\quad n_2\in\mathbb{Z},\quad k-n_1-n_2\ge0 ,
    \label{eq:regimeIII}
\end{align}
i.e., products of nonnegative powers of $C/|\Sr|^{2/3}$, $\Sp/|\Sr|^{2/3}$, and
$|\Sr|^{2/3}$.

The elimination is that of basis II with the other constant in the role of the
label. Equation \eqref{eq:regimeIII} again leaves no power of $C$ to eliminate:
the term $k=n_1=n_2=0$ carries none, so every observable tends to a
$C$-independent limit fixed by the segment constant alone,
\begin{equation}
    \cU\;\longrightarrow\;\cO^{\cU}_{(0;0,0)}(\gamma)\,|\Sr|^{2/3}
    \qquad\text{as}\quad C/|\Sr|^{2/3}\to0 .
    \label{eq:interceptIII}
\end{equation}
The inversion must therefore be carried out around that limit. The
shifted variable
\begin{align}
    \widehat{\cU}_b&\equiv\frac{\cU_b}{|\Sr|^{2/3}}-\cO^{\cU_b}_{(0;0,0)}
    \nonumber\\
    &=\cO^{\cU_b}_{(1;0,0)}\,\frac{C}{|\Sr|^{2/3}}
    \Bigl[1+O\bigl(C/|\Sr|^{2/3}\bigr)\Bigr]
    \label{eq:shiftedIII}
\end{align}
vanishes linearly in $C$, so Lagrange inversion applies to it in place of
Eq.~\eqref{eq:xseries}. Substituting the resulting $C(\widehat{\cU}_b)$ into
$\cU_a$ gives
\begin{align}
    \log\cU_a&=\frac23\log|\Sr|
    \nonumber\\
    &\quad+\sum_{\substack{k,n_1\in\mathbb{N}_0,\ n_2\in\mathbb{Z}\\ k-n_1-n_2\ge0}}
    \cM^{\log}_{(k;n_1,n_2)}(\gamma)\,
    \Sp^{\,n_1}\,|\Sr|^{\frac23 n_2}
    \nonumber\\
    &\qquad\times
    \widehat{\cU}_b^{\;k-n_1-n_2} ,
    \label{eq:relIII}
\end{align}
where the $(0;0,0)$ term is $\log\cO^{\cU_a}_{(0;0,0)}$. Again no term proportional
to $\log\cU_b$ survives.

\medskip\noindent
\textbf{Basis IV.} 
For ordering (F), in which $|\sigma_r|\gg1$ with
$|\Sr|^{2/3}\ll\Sp$, the seed is that of basis III and the surface
condition that of basis II, and the monomials mix the two:
\begin{align}
    \text{IV}:\quad
    &\Sp^{\,n_1}\,|\Sr|^{\frac23 n_2}\,C^{\,k+|n_1|-n_2},
    \nonumber\\
    &n_1,n_2\in\mathbb{Z},\quad k\in\mathbb{N}_0,\quad k+|n_1|-n_2\ge0 .
    \label{eq:regimeIV}
\end{align}
The $C$-independent limit is again $\cO_{(0;0,0)}\Sp$, and the
inversion runs on the shifted variable \eqref{eq:shiftedII}, with $|\Sr|^{2/3}$
entering the coefficients. The master relation is
\begin{align}
    \log\cU_a&=\log\Sp+\log\cM_{(0;0,0)}(\gamma)
    \nonumber\\
    &\quad+\!\!\sum_{\substack{n_2\in\mathbb{Z},\ k\in\mathbb{N}\\ k-n_2\ge0}}\!\!
    \cM^{\log}_{(k;0,n_2)}(\gamma)\,|\Sr|^{\frac23 n_2}\,
    \widetilde{\cU}_b^{\;k-n_2}
    \nonumber\\
    &\quad+\!\!\sum_{\substack{n_1\in\mathbb{Z}\backslash\{0\},\ n_2\in\mathbb{Z},
    \ k\in\mathbb{N}_0\\ k+|n_1|-n_2\ge0}}\!\!
    \cM^{\log}_{(k;n_1,n_2)}(\gamma)\,\Sp^{\,n_1}\,|\Sr|^{\frac23 n_2}
    \nonumber\\
    &\qquad\times
    \widetilde{\cU}_b^{\;k+|n_1|-n_2} .
    \label{eq:relIV}
\end{align}
Again, no term proportional to $\log\cU_b$ appears, so that the relation is once more a
$C$-independent limit fixed by the segment constants together with a series
in the shifted variable.

\section{Dimensionless form of the single-polytrope system}
\label{app:LE}

\begin{table*}[t]
\centering
\caption{Parameters of the fit \eqref{eq:fitform} to the coefficients
$M_k(\gamma)$ of $(\Ib,\lbt)$.}
\label{tab:fitI}
\begin{tabular}{l|rrrrrr}
\hline\hline
 & $a_{k0}$ & $a_{k1}$ & $a_{k2}$ & $a_{k3}$ & $a_{k4}$ & $a_{k5}$ \\
\hline
$M_0$ & $-0.63886$ & $-0.01115$ & $+0.16285$ & $-0.83254$ & $+1.85744$ & $-1.56211$ \\
$M_1$ & $+2.73639$ & $-0.03459$ & $+0.29315$ & $-1.81473$ & $+3.87202$ & $-3.27567$ \\
$M_2$ & $-0.74339$ & $-0.02585$ & $+0.34795$ & $-1.61873$ & $+3.54061$ & $-2.60995$ \\
$M_3$ & $+0.17320$ & $-0.02536$ & $+0.47673$ & $-2.43398$ & $+5.16068$ & $-3.94469$ \\
\hline\hline
\end{tabular}
\end{table*}

\begin{table*}[t]
\centering
\caption{Parameters of the fit \eqref{eq:fitform} to the coefficients
$M_k(\gamma)$ of $(\Qb,\lbt)$.}
\label{tab:fitQ}
\begin{tabular}{l|rrrrrr}
\hline\hline
 & $a_{k0}$ & $a_{k1}$ & $a_{k2}$ & $a_{k3}$ & $a_{k4}$ & $a_{k5}$ \\
\hline
$M_0$ & $+1.27781$ & $+0.02107$ & $-0.32652$ & $+1.70702$ & $-3.84146$ & $+3.23161$ \\
$M_1$ & $-3.13441$ & $+0.08165$ & $-0.42523$ & $+2.89997$ & $-6.18910$ & $+5.29540$ \\
$M_2$ & $+1.57231$ & $+0.04673$ & $-0.55072$ & $+2.30153$ & $-4.32301$ & $+2.55783$ \\
$M_3$ & $+0.94798$ & $-0.57632$ & $+6.00178$ & $-25.02400$ & $+42.73209$ & $-26.68657$ \\
\hline\hline
\end{tabular}
\end{table*}

\begin{table*}[t]
\centering
\caption{Parameters of the fit \eqref{eq:fitform} to the coefficients
$M_k^{(QI)}(\gamma)$ of the $\Qb$--$\Ib$ relation \eqref{eq:IQrel}.}
\label{tab:fitIQ}
\begin{tabular}{l|rrrrrr}
\hline\hline
 & $a_{k0}$ & $a_{k1}$ & $a_{k2}$ & $a_{k3}$ & $a_{k4}$ & $a_{k5}$ \\
\hline
$M_0$ & $+1.59724$ & $+0.02669$ & $-0.40781$ & $+2.12091$ & $-4.76349$ & $+4.00714$ \\
$M_1$ & $-3.27141$ & $+0.08821$ & $-0.65197$ & $+3.96712$ & $-8.58710$ & $+7.31764$ \\
$M_2$ & $-2.22572$ & $+0.15913$ & $-1.36404$ & $+7.64869$ & $-16.17694$ & $+13.12886$ \\
$M_3$ & $-1.83626$ & $+0.06070$ & $-0.25287$ & $+4.36042$ & $-13.04939$ & $+13.23853$ \\
\hline\hline
\end{tabular}
\end{table*}

We present the dimensionless system of equations for the single-polytrope stars used in
Sec.~\ref{sec:gamma}. 
We define the dimensionless pressure and energy density as 
\begin{equation}
    \theta\equiv\frac{p/\epsilon}{q_c}=\left(\frac{p}{p_c}\right)^{1-\gamma},
    \qquad
    \frac{p}{p_c}=\theta^{\gp},
    \qquad
    \frac{\epsilon}{\ec}=\theta^{\gm}.
    \label{eq:LEtheta}
\end{equation}
Here, $\theta$ equals $1$ at the center and $0$ at the surface. 
With the length
and mass variables defined as
\begin{align}
    &\xi\equiv\frac{r}{a},\qquad
    a^2\equiv\frac{p_c}{4\pi\,(1-\gamma)\,\ec^{\,2}},
    \qquad
    \mu\equiv\frac{m}{4\pi a^{3}\ec},\\
    &u\equiv\frac{q_c}{1-\gamma},\qquad
    c=\frac{u\,\mu}{\xi},
    \label{eq:LExi}
\end{align}
the constant $K$ of Eq.~\eqref{eq:eos} drops out of the system, and the only
remaining parameters are $\gamma$ and $q_c$. Equation \eqref{eq:TOV}
becomes
\begin{align}
    &\frac{d\mu}{d\xi}=\xi^{2}\,\theta^{\gm},
    \qquad L=1-2c,
    \nonumber\\
    &\frac{d\theta}{d\xi}
    =-\frac{\bigl(1+q_c\theta\bigr)\bigl[\mu+q_c\,\xi^{3}\,\theta^{\gp}\bigr]}
           {\xi^{2}\,L},
    \label{eq:LEbg}
\end{align}
which reduces to the Lane--Emden equation as $q_c\to0$. Equation \eqref{eq:omegadiff} becomes, with
$w\equiv\omega_1/\omega_c$ and
$V\equiv u^{-1}dw/d\xi$,
\begin{align}
    \frac{dw}{d\xi}&=u\,V,
    \nonumber\\
    \frac{dV}{d\xi}&=-\frac{4}{\xi}
    \left[1-\frac{u\,\xi^{2}\,\theta^{\gm}\bigl(1+q_c\theta\bigr)}{4L}\right]V
    \nonumber\\
    &\quad+\frac{4\,\theta^{\gm}\bigl(1+q_c\theta\bigr)}{L}\,w ,
    \label{eq:LEomega}
\end{align}
and Eq.~\eqref{eq:H0diff} becomes, with $H\equiv H_0$ and $G\equiv dH/d\xi$,
\begin{align}
    \frac{dH}{d\xi}&=G,
    \nonumber\\
    \frac{dG}{d\xi}&=-\,\fa\,G-\fb\,H
    -\frac{\gm\,\bigl(1+q_c\theta\bigr)\,\theta^{\gm}}{L\,\theta}\,H,
    \label{eq:LEtidal}
\end{align}
where
\begin{align}
    \fa&=\frac{1}{\xi}\left[1+\frac{1-u\,\xi^{2}
      \bigl(\theta^{\gm}-q_c\theta^{\gp}\bigr)}{L}\right],\nonumber\\
    \fb&=-\frac{6}{L\,\xi^{2}}
      +\frac{u\bigl(5\,\theta^{\gm}+9\,q_c\,\theta^{\gp}\bigr)}{L}
    \nonumber\\
    &\quad-\left[\frac{2u\bigl(\mu+q_c\xi^{3}\theta^{\gp}\bigr)}{\xi^{2}L}\right]^{2} .
    \label{eq:LEab}
\end{align}
The last term of Eq.~\eqref{eq:LEtidal} originates from $d\epsilon/dp$, which
diverges at the surface: for $\gamma<1/2$ this term diverges at the surface, since
$\theta^{\gm}/\theta\to\infty$ as $\theta\to0$. 
If we set
\begin{equation}
    \Psi\equiv\frac{1+q_c\theta}{L\,\bigl(d\theta/d\xi\bigr)},
    \qquad
    \hat G\equiv G+\theta^{\gm}\,\Psi\,H,
    \label{eq:LEGhat}
\end{equation}
the variable $\hat G$ is smooth up to the surface, with
\begin{align}
    \frac{d\hat G}{d\xi}
    &=-\,\fa\,G-\fb\,H
    +\theta^{\gm}\bigl[\Psi'\,H+\Psi\,G\bigr],
    \nonumber\\
    G&=\hat G-\theta^{\gm}\Psi H .
    \label{eq:LEGhateq}
\end{align}
The integrated variables are the six components $(\xi,\mu,w,V,H,\hat G)$.

The surface $\theta\to0$ is a finite point in $\xi$, but the exponent of 
$\theta^{\gamma/(1-\gamma)}$ is generally nonintegers, so we take
$t\equiv-\log\theta$ as independent variable:
\begin{equation}
    \frac{dY}{dt}=\frac{dY}{d\xi}\cdot\frac{d\xi}{dt},
    \qquad
    \frac{d\xi}{dt}=-\frac{\theta}{\;d\theta/d\xi\;} .
    \label{eq:LEt}
\end{equation}
The initial conditions at the center follow from the expansion of
Eq.~\eqref{eq:LEbg},
$\theta=1-\frac{(1+q_c)(1+3q_c)}{6}\xi^{2}+O(\xi^4)$, $\mu=\xi^3/3+O(\xi^5)$,
and Eq.~\eqref{eq:centralpert}: with $\theta_0=1-\delta$ ($\delta\ll1$),
\begin{align}
    &\xi_0=\left[\frac{6\,\delta}{(1+q_c)(1+3q_c)}\right]^{1/2},
    \qquad
    \mu_0=\frac{\theta_0^{\gm}\,\xi_0^{3}}{3},
    \nonumber\\
    &w_0=1,\qquad V_0=\frac{4}{5}\bigl(1+q_c\bigr)\xi_0,
    \nonumber\\
    &H(\xi_0)=\xi_0^{2},\qquad G(\xi_0)=2\xi_0 .
    \label{eq:LEinit}
\end{align}
The overall normalizations of $w$ and $H$ cancel in Eq.~\eqref{eq:observables}.

\subsection{Extraction of the coefficients \texorpdfstring{$M_k$}{Mk}}
\label{app:Mk}

The integration of Sec.~\ref{sec:PN} returns the two surface series
\begin{equation}
    q_c^{2}\Ib=\sum_{j\ge0}\cI_j(\gamma)\,q_c^{\,j},
    \qquad
    q_c^{5}\lbt=\sum_{j\ge0}\cL_j(\gamma)\,q_c^{\,j},
    \label{eq:ILseries}
\end{equation}
both finite at $q_c=0$. Equation~\eqref{eq:fdef} and the definition of the
expansion variable are then
\begin{align}
    F(q_c)&\equiv\log\Bigl(\sum_j\cI_jq_c^{\,j}\Bigr)
    -\frac25\log\Bigl(\sum_j\cL_jq_c^{\,j}\Bigr)
    \nonumber\\
    &=\sum_{j\ge0}f_j\,q_c^{\,j},\label{eq:FXdef}
\end{align}
and
\begin{align}
    X&\equiv\lbt^{-1/5}
    =q_c\sum_{j\ge0}x_j\,q_c^{\,j},\label{eq:Xdef}
\end{align}
with $x_j$ the coefficients of $(\sum_j\cL_jq_c^{\,j})^{-1/5}$. Since
$x_0\neq0$, Eq.~\eqref{eq:Xdef} is inverted to obtain
$q_c(X)$, and substituting it into $F$ expresses
$\log\Ib-\frac25\log\lbt=F$ as a series in $X$, which is
Eq.~\eqref{eq:mainrelation}: $M_k$ is the $k$-th Taylor coefficient of
$F\bigl(q_c(X)\bigr)$ in $X$. Writing the reversion
as $q_c=\sum_{k\ge1}A_kX^{k}$ with
\begin{equation}
    A_1=\frac{1}{x_0},
    \qquad
    A_2=-\frac{x_1}{x_0^{3}},
    \qquad
    A_3=\frac{2x_1^{2}-x_0x_2}{x_0^{5}},
    \qquad \dots
    \label{eq:reversion}
\end{equation}
one obtains
\begin{align}
    &M_0=f_0,
    \qquad
    M_1=\frac{f_1}{x_0},
    \qquad
    M_2=\frac{f_2}{x_0^{2}}-\frac{f_1x_1}{x_0^{3}},
    \nonumber\\
    &M_3=\frac{f_3}{x_0^{3}}-\frac{2f_2x_1}{x_0^{4}}
    +\frac{f_1\bigl(2x_1^{2}-x_0x_2\bigr)}{x_0^{5}},
    \label{eq:Mkexplicit}
\end{align}
and so on. The $k=0$ term reproduces Eq.~\eqref{eq:M0}, since
$f_0=\log i-\frac25\log(\frac23k_2^{\rm N})$ by Eq.~\eqref{eq:Inewt}. Each $f_j$ and $x_j$ is obtained from a quadrature of the $j$-th-order system in Eq.~\eqref{eq:PNexp}, so
the coefficient $M_k$ at each post-Newtonian order is obtained in closed form once that
system has been integrated.

\section{Analytic expressions for the coefficient functions}
\label{app:fits}

For the observable pairs $(\cU_a,\cU_b)=(\Ib,\lbt)$ and $(\Qb,\lbt)$ we provide analytic fitting formulas for the coefficient functions of the relation
\begin{equation}
    \log\cU_a=\frac{\delta_a}{\delta_b}\log\cU_b
    +\sum_{k\geq0}M_k^{(\cU_a\cU_b)}(\gamma)\,x^{k},
    \qquad x\equiv\lbt^{-1/5}.
    \label{eq:appseries}
\end{equation}
According to Eq.~\eqref{eq:deltas}, the slope is $2/5$ for
$(\Ib,\lbt)$ and $1/5$ for $(\Qb,\lbt)$. The error of the truncation at
$k\le3$, measured by the size of the dropped $k=4$--$6$ terms, does not exceed
$4\times10^{-3}$ for $\lbt\ge30$ and $1.4\times10^{-3}$ for $\lbt\ge10^{2}$.

The $(\Ib,\lbt)$ coefficients are the quadrature values of Sec.~\ref{sec:VC}.
For $(\Qb,\lbt)$, the coefficients were extracted from sequences obtained by
integrating the rotational quadrupole sector
\eqref{eq:dk2dc}--\eqref{eq:mu2} at finite $q_c$. The
Newtonian limit was pinned by the closed form
\begin{equation}
    M_0^{(Q\lambda)}=\frac45\log\Bigl[\frac23\,k_2^{\rm N}(\gamma)\Bigr]
    -2\log i(\gamma) .
    \label{eq:M0Q}
\end{equation}
The systematic error of this extraction procedure was calibrated by applying the same
procedure to $(\Ib,\lbt)$ and comparing with the quadrature values: for
$\gamma\le0.64$ it is at the level of $10^{-6}$ for $M_1$, $10^{-4}$ for $M_2$,
and $10^{-2}$ for $M_3$.

The fitting form is a common quintic polynomial for all coefficients,
\begin{equation}
    M_k^{(\cU_a\cU_b)}(\gamma)=\sum_{j=0}^{5}a_{kj}\,\gamma^{j},
    \qquad 0\le\gamma\le0.64,
    \label{eq:fitform}
\end{equation}
with the parameters $a_{kj}$ listed in Tables~\ref{tab:fitI} and
\ref{tab:fitQ}. The maximum relative difference between the fits and the
numerical values is below $0.15\,\%$ for $M_0$--$M_2$ and $1.1\,\%$ for $M_3$
in the $(\Ib,\lbt)$ case, and below $0.07\,\%$ for $M_0$--$M_2$ and
$2.5\,\%$ for $M_3$ in the $(\Qb,\lbt)$ case (the $M_3$ differences are
comparable to the numerical extraction uncertainty quoted above).

The $(\Qb,\Ib)$ relation is obtained in the same way, by setting
$x=\Ib^{-1/2}$ in Eq.~\eqref{eq:appseries}:
\begin{equation}
    \log\Qb=\frac12\log\Ib+\sum_{k\ge0}M_k^{(QI)}(\gamma)\,\Ib^{-k/2} .
    \label{eq:IQrel}
\end{equation}
The fit \eqref{eq:fitform} to its coefficients is listed in
Table~\ref{tab:fitIQ}; truncated at $k\le3$, Eq.~\eqref{eq:IQrel} is accurate
to $1.5\times10^{-2}$ for $\lbt\ge30$ and $6.7\times10^{-3}$ for
$\lbt\ge10^{3}$.

\bibliography{refs}

\end{document}